\documentclass[a4paper,11pt]{article}

\usepackage{graphics,color,array,dcolumn}
\usepackage{graphicx}
\usepackage{amssymb}
\usepackage{xspace}
\usepackage{color}

\usepackage{amsfonts}
\usepackage{latexsym}
\usepackage{amsmath}
\usepackage{vmargin}

   \def\be{\begin{equation}}
   \def\ee{\end{equation}}
   \def\ba{\begin{eqnarray}}
   \def\ea{\end{eqnarray}}

   \def\half{{1\over 2}}

   \newcommand{\fss}[1]{#1\!\!\!/}
   \newcommand{\cin}{\fss{\partial}}

   \newcommand{\bq}{{\bar{q}}}
   \newcommand{\bp}{{\bar{p}}}
   \newcommand{\bphi}{{\bar{\phi}}}
   \newcommand{\bpsi}{{\bar{\psi}}}
   \newcommand{\psid}{{\psi^\dagger}}
   \newcommand{\bpi}{{\bar{\pi}}}

   \newcommand{\bth}{{\bar{\theta}}}
   \newcommand{\dt}{{\partial_t}}
   \newcommand{\id}{{\, i \dt}} 
   \newcommand{\de}{{\delta}}
   \newcommand{\pa}{{\partial}}
   \newcommand{\cH}{{\cal H}}

\setmarginsrb{30mm}{20mm}{30mm}{20mm}%
             {0mm}{10mm}{0mm}{10mm}
\title{Functional RG flow of the effective Hamiltonian action}
\author{G. P. Vacca\thanks{vacca@bo.infn.it} \ and 
        L. Zambelli\thanks{Luca.Zambelli@bo.infn.it}\\ 
\\
Dip. di Fisica, Universit\`a degli Studi di Bologna\\
INFN Sez. di Bologna\\
via Irnerio 46,  I-40126 Bologna,  Italy}
\date{}

\begin{document}
\maketitle

\begin{abstract}
After a brief review of the definition and properties of the quantum effective Hamiltonian action we describe its renormalization flow by a functional RG equation.
This equation can be used for a non-perturbative quantization and study also of theories with bare Hamiltonians which are not quadratic in the momenta. 
As an example the vacuum energy and gap of quantum mechanical models are computed. 
Extensions of this framework to quantum field theories are discussed. In particular one possible Lorentz covariant approach for simple scalar field theories is developed.
Fermionic degrees of freedom, being naturally described by a first order formulation, can be easily accommodated in this approach.


\end{abstract}

\section{Introduction}
Quantum mechanical systems can be studied with a variety of methods such as for example the canonical operatorial approach or the functional methods.
The latter are usually employed to construct generating functionals of different kinds of correlators from which the physical observables of interest can be derived.
Among them a very useful object is the so called quantum effective action, a functional mostly used in quantum field theory (QFT), in both perturbative and non perturbative approaches.
This is in general a highly non local object which encodes all the quantum properties of the system; for instance, it generates the proper vertices of the theory. 

The effective action most commonly discussed in the literature
is of the Lagrangian type, since it is derived from the second order Lagrangian formulation of the bare theory.
There is a very good reason to do that, namely that people usually consider bare Hamiltonians which are quadratic in the momenta such that one can easily move to a Lagrangian description. The rationale for
this is obtaining a manifestly Lorentz-covariant formulation in $d$ space-time dimensions. Another advantage of passing to a second order formulation is that the number of
fields in configuration space is half the one in phase space, since in the functional formulation the conjugated momenta have been integrated out. 
 
On the other hand one may also consider the reasons to choose a first order Hamiltonian description on the phase space of a theory.
Clearly this is unavoidable when dealing with the quantization of theories with bare Hamiltonians non quadratic in the momenta. In such a case the full phase space variables
are needed for a quantum description of the system.
Traditionally the main advantage attributed to the 
Hamiltonian formulation is that it makes unitarity manifest~\cite{Weinberg1}. This is due to the strict relationship established
by canonical quantization between the
classical symplectic structure on phase space and the inner product on
the Hilbert space.
The Hamiltonian approach may be useful also when configuration
space is not a vector space, since phase space can usually be interpreted as a cotangent bundle and it could be easier to deal with. In the functional integral representation this is translated in the possibility that the measure in phase space be field independent while the one in configuration space be not. This happens for instance in the case of non linear sigma models. Of course, even in this case 
whenever the bare theory is quadratic in the momenta the Lagrangian and the Hamiltonian formulations lead to the same results (Matthews Theorem), as proved by perturbative studies~\cite{Matthews,BN}.
In a functional integral representation, the Hamiltonian approach is based on quantum generating functionals obtained introducing sources in the phase space path integral~\cite{sakita}.
From them, one can define a quantum effective Hamiltonian action which generates the proper vertices. This was recently studied in~\cite{VCAS},
on the wake of a renewed interest in Hamiltonian gauge theories such as QCD, in particular in the Coulomb gauge (see~\cite{ReinhardtWatson} and references therein).

The purpose of the present work is to present a non-perturbative framework which allows to compute, within specific approximation schemes, the quantum effective Hamiltonian action.
This approach is based on the definition of a one-parameter-family of deformed effective actions, which were introduced in the literature
a long time ago under the name of average effective (Lagrangian) actions~\cite{wetterich}. This is only one of the many formulations of the
functional renormalization group (RG)~\cite{Wilson, WH,Pol},
which is a Wilsonian representation of QFT based on a coarse-graining procedure allowing to interpolate, 
by moving along an RG flow trajectory in theory
space, between the bare theory at the ultraviolet (UV) scale and the quantum effective theory at the infrared (IR) scale. 
Providing a bare action in the UV and solving the
RG flow equation with an appropriate set of boundary conditions, one could in principle obtain the quantum effective action.
Since this equation is a functional differential equation, generally one
is forced to employ specific approximation schemes, essentially strongly constraining the space of functionals the solution belongs to
with the help of physical arguments.
For almost two decades such a theoretical framework has been applied to investigate several aspects of QFT's and
condensed matter systems~\cite{BTW} and has ben used to probe the possibility that for example Einstein gravity, as a QFT,
can be nonperturbatively renormalizable~\cite{Reuter} within the paradigm of asymptotic safety~\cite{WeinbergAS}.

In a previous work~\cite{VZ} we proposed the use of
cutoff operators affecting the symplectic form of phase space
 and implementing
a more balanced coarse-graining and regularization, with respect to the cases where the coarse-graining is performed on the fluctuations in configuration space only, but after this choice of regularization, 
we restricted our discussion to bare Hamiltonians quadratic in the momenta and we fully integrated out the momenta, obtaining a cutoff dependent functional measure in the Lagrangian path integral, which was leading to a subtraction term in the RG flow equation. 
Here instead we are interested in retaining the full dynamics in phase space, building a flow which realizes the idea of shell-by-shell simultaneous integration on both phase space variables.
As a disclaimer let us add that other non-perturbative RG
flows called ``Hamiltonian flows" already appeared in the literature, but they largely differ from our formulation. Examples are the similarity RG~\cite{WG}, which is
generated by iterated unitary transformations within the operatorial representation,
and the flows based on a variational solution of the Schr\"odinger 
equation~\cite{RP}.

In this paper we start our discussion from quantum mechanical systems
($0\!+\!1$  dimensional QFT's) with scalar degrees of freedom, for which we review some of the properties of the Hamiltonian effective action in the first part of Section $2$, and we prove some formula useful for the subsequent developments. 
In the second part of Section $2$ we derive the main equations  
satisfied by the average effective Hamiltonian action (AEHA) of a quantum mechanical system. They depend on a cutoff operator which suppresses part of the functional integration
generating a one-parameter flow from the UV to the IR. 
In particular we give the simpler equations associated to the so called local Hamiltonian approximation (LHA), which is the lowest order term of the derivative expansion of the full functional, for some specific cutoff operators. 
These are then used (Section $2.3$) to study a family of exactly solvable Hamiltonians which are not quadratic in the momenta and indeed we show that one can easily extract informations like the ground state and the first energy gap of such systems. The same approach can be used to study general systems with arbitrary bare Hamiltonians.
We conclude Section $2$ discussing the extension of the formalism to quantum mechanical theories with fermionic degrees of freedom.

In section $3$ we start to address quantum field theories. The extension to the non covariant version of QFT is straightforward and we first discuss it briefly for the case of scalar QFT.
Since in the traditional Hamiltonian formulation of QFT one pays explicit
unitarity with the disguising of Lorentz invariance, we discuss one
possible way around this drawback, that is, we spend the last part of the paper in discussing a manifestly Lorentz symmetric 
(but maybe not manifestly unitary) extension of 
the previous framework inside the realm of the covariant Hamiltonian formalism. 

This is a subject which has a long history in classical physics
~\cite{DW,GMS,GIMM},
but whose applications to quantum dynamics are pretty rare to be found in the literature.
Even if under different names, the covariant Hamiltonian formulation
of Yang Mills theory is one of the oldest examples.
M.B. Halpern in 1977 addressed such a formalism for QCD, generically naming it ``first order formalism''~\cite{Halpern}
but he immediately abandoned the full phase space formulation
integrating out the gauge vector fields
thus being left with a theory, containing only conjugate momenta,
that he called ``field strenght formulation'', 
which was studied in the following years 
(see~\cite{LR} and references in it).
More recently a slight variant of the ``first order formalism" (still covariant) for Yang Mills theory has received fresh attentions 
from the perspective of topological BF theories~\cite{CCR}.
In particular the reader can find in~\cite{ZM} an explicit one
loop computation of what we call the effective covariant Hamiltonian
action of pure Yang Mills, reproducing the expected asymptotic freedom result.
Despite these successful examples, the main open question about
covariant Hamiltonian QFT is still about
its foundations, even if these have begun to be studied recently by some author~\cite{Kanatchikov,BS}. 
These investigations can shed light on the issue of unitarity
of this covariant formulation.
Without a sound Lorentz covariant
quantization prescription, covariant Hamiltonian formalism seems but 
a game, legitimate only in the special case of Hamiltonians quadratic
in the momenta. On the other hand, only by studying this approach in more general cases and by looking for its applications to real physical systems one can hope to find a legitimation for the search of foundations.
 
In this work, for what concerns a covariant Hamiltonian formulation of QFT's, we restrict ourselves to defining the average effective covariant Hamiltonian action of a scalar field theory in a particularly simple case.
This consists in assuming that the non trivial dependence on the covariant momenta is in the longitudinal (w.r.t. Fourier variable) subspace of the space of conjugate momenta.
This definition is compatible with both QM in $0\!+\!1$ dimensions 
and with QFT's whose bare Hamiltonians are quadratic in the momenta, 
and it provides a particular dynamical extension outside this domain.
For this simple case we present a framework for studying such a model 
by a non-perturbative RG flow equation.
For completeness we also comment on the corresponding covariant Hamiltonian formulation for theories with Dirac fermions.

In the conclusions the reader will find a discussion about the physical
motivations for the introduction of this formalism, as well as
a proposal of some possible developments, extensions and future
applications of this method.
Several appendices follow, where some technical issues are described in more details.

\section{The effective Hamiltonian action in quantum mechanics}
In this section we shall work within quantum mechanics (QM), i.e. a 
$0\!+\!1$ dimensional quantum field theory (QFT).
As an example we will quantize a classical system with one bosonic degree of freedom
governed by the following Hamiltonian action:
\be\label{QMhamiltonianaction}
S[p,q]=\int\!\! dt\, \Big[\, p(t)\dt q(t) - H\left( p(t),q(t) \right) \Big]
\ee
where the (bare) Hamiltonian can have an arbitrary dependence in the momenta, departing therefore from the usual quadratic form 
\be\label{QMbarehamiltonian}
 H(p,q) = \frac{1}{2}\, p^2 + V(q)\,.
 \ee
Here and in the following $p$ and $q$ denote canonically conjugate variables.
The quantization of such a system is performed via the following 
phase-space path integral: 
\be\label{hampathint1}
e^{\frac{i}{\hbar} W[I,J]}=\int \left[ dp dq\right]\mu[p,q] 
e^{\, \frac{i}{\hbar}\left\{S[p,q] + I\cdot p + J\cdot q  \right\} }
\ee
where the dots stand for ordinary integrations.
The functional measure on the physical phase space is usually assumed to be
$\mu[p,q]={\rm Det}\frac{1}{2\pi\hbar}$.Also one can easily extend all the formalism to an
euclidean description.
Since we want to keep our discussion as general as possible we will not specify 
the precise space of functions on which the functional integral is defined. 

It is possible to study the system by a functional which may be called
the quantum effective Hamiltonian action, which is a trivial generalization of the
more widely known effective Lagrangian action.
The latter $\Gamma^L$ is defined by introducing in the
configuration-space path integral 
external sources $J$ coupled to the Lagrangian variables,
and by taking the Legendre transform of the generating functional 
of the connected green's functions $W[J]$
 with respect to (w.r.t.) $J$.
Similarly, in order to define the effective Hamiltonian action $\Gamma^H$,
one starts from the phase-space path integral~\eqref{hampathint1}
 and performs a Legendre transform:
\be\label{legendre}
\Gamma^H\left[\bp,\bq\right] =
\mathop{\rm ext}_{I,J}\left(W[I,J] -I\cdot \bp - J \cdot\bq \right)\,,
\ee
where
\be
\bp=\frac{\delta W}{\delta I}  \quad , \quad 
\bq=\frac{\delta W}{\delta J}\, .  \nonumber
\ee
The introduction of such a functional is not a novelty, as we have discussed in the introduction.
There are several ways to convince ourselves that from this functional 
one can get every information about the quantum system.

\noindent
{\it First}, by taking functional derivatives w.r.t. $\bq(t)$ and $\bp(t)$ one immediately
gets
\be\label{eqIJ}
I=-\frac{\delta \Gamma^H}{\delta \bp} \quad , \quad
J=-\frac{\delta \Gamma^H}{\delta \bq}\,\,. 
\ee
For zero sources one has the equations for the vacuum configuration
$(\bq,\bp)$. They appear as the classical equations of motion
obtained from the quantum effective Hamiltonian action.

\noindent
{\it Second}, $\Gamma^H$ satisfies the following integro-differential equation
\be\label{intdiffeq}
e^{\frac{i}{\hbar}\Gamma^H[\bp,\bq]}=\!\int\! \left[ dp dq\right]\mu[p,q] 
e^{\, \frac{i}{\hbar} \left\{ S[p,q]
-(q-\bq)\cdot\frac{\delta \Gamma^H}{\delta \bq}-(p-\bp)\cdot\frac{\delta \Gamma^H}{\delta \bp}
\right\} }\,.
\ee
This is a central identity and it could also be promoted to the definition of $\Gamma^H$.

\noindent
{\it Third}, from this equation one can get a different proof that the classical equations
satisfied by the effective Hamiltonian action encode the full quantum dynamics, 
because they are equivalent to the Hamiltonian Dyson-Schwinger equations. 
In fact, the identities: 
\ba
0
&=&\!\int\! \left[ dp dq\right]\frac{\delta}{\delta p}\left(
\mu[p,q] e^{\, \frac{i}{\hbar} \left\{ S[p,q]
-(q-\bq)\cdot\frac{\delta \Gamma^H}{\delta \bq}-(p-\bp)\cdot\frac{\delta \Gamma^H}{\delta \bp}
\right\} }
\right) \nonumber \\
&=&\!\int\! \left[ dp dq\right]\frac{\delta}{\delta q}\left(
\mu[p,q] e^{\, \frac{i}{\hbar} \left\{ S[p,q]
-(q-\bq)\cdot\frac{\delta \Gamma^H}{\delta \bq}-(p-\bp)\cdot\frac{\delta \Gamma^H}{\delta \bp}
\right\} } \right) \nonumber 
\ea
lead to:
\be
\langle
-i\hbar\frac{\delta}{\delta p}\log \mu[p,q]+\frac{\de S}{\de p}
\rangle
=\frac{\delta \Gamma^H}{\delta \bp} \quad , \quad
\langle
-i\hbar\frac{\delta}{\delta q}\log \mu[p,q]+\frac{\de S}{\de q}
\rangle
=\frac{\delta \Gamma^H}{\delta \bq} \,. \nonumber 
\ee

\noindent
{\it Forth}, just like for the effective action,
the effective Hamiltonian action has a similar interpretation as the generator 
of the one-particle-irreducible (1PI) proper vertices. 
For more details and a proof of this statement see Appendix A.

\noindent
{\it Fifth}, by evaluating the effective Hamiltonian action on its
stationarity $\bp$ values one gets the effective Lagrangian action.
In fact, defining
\be
\Gamma^L[\bq]=\mathop{\rm ext}_{\bp}\Gamma^H[\bp,\bq]\nonumber
\ee
and calling $\bp_\bq$ the extremal point, it is straightforward to show that
\be
I=-\frac{\delta \Gamma^H}{\delta \bp}[\bp_\bq,\bq]=0  \quad , \quad 
J=-\frac{\delta \Gamma^H}{\delta \bq}[\bp_\bq,\bq]=-\frac{\de\Gamma^L}{\de \bq}[\bq] \,. \nonumber
\ee
Therefore
$\Gamma^L[\bq]= W\left[ 0,-\frac{\de\Gamma^L}{\de \bq} \right] +\bq\cdot \frac{\de\Gamma^L}{\de \bq}\, $,
wherefrom we learn that $\Gamma^L$ satisfies the integro-differential equation:
\be
e^{\frac{i}{\hbar}\Gamma^L[\bq]}=\!\int\! \left[dp dq\right]\mu[p,q] 
e^{\, \frac{i}{\hbar} \left\{S[p,q]- (q-\bq)\cdot\frac{\de \Gamma^L}{\de \bq}\right\}}\nonumber
\ee
which is a generalization of the usual configuration space integro-differential equation satisfied by the effective action,
since it does not require $S$ to be quadratic in the momenta. 
Due to this simple relation between the two effective actions, from here on
and for the rest of this paper we will use the same letter $\Gamma$ for both,
dropping the subscripts, since the reader will be able to distinguish them by their
arguments ($\bp$,$\bq$ for the Hamiltonian one and $\bq$ only for the Lagrangian one).

\noindent
{\it Sixth}, the effective Hamiltonian action can be defined from the operatorial representation
by means of a time-dependent variational principle, in a way which is the direct generalization of the usual construction in configuration space~\cite{JK}.
Let $\hat H$ be the Hamiltonian operator of the quantum system, 
$|0\rangle$ be its time-independent ground state
and let the boundary conditions of the path integral in~\eqref{hampathint1} be chosen such that
\be\label{vacuum W}
e^{\frac{i}{\hbar}W[I,J]}=\langle0|\hat U_{I,J}(+\infty,-\infty)|0\rangle=
\langle0|T\exp\left\{-\frac{i}{\hbar}\int_{-\infty}^{+\infty}\!\!dt\left[\hat H-J(t)\hat q-I(t)\hat p\right]\right\}|0\rangle\, .
\ee
Then the effective Hamiltonian action defined in~\eqref{legendre} is related in the following way
\be\label{variational def}
\Gamma\left[\bp,\bq\right] =
\mathop{{\rm ext}}_{|\psi_\pm, t\rangle}\left(\int_{-\infty}^{+\infty}\!\! dt\, \langle \psi_- ,t|i\hbar\dt-\hat H| \psi_+ ,t\rangle \right)
\ee
to an extremum with respect to variations of the two states $|\psi_{\pm}, t\rangle$ preserving the constraints
\be\label{constraints psi}
\langle \psi_- ,t| \psi_+ ,t\rangle=1 \quad , \quad 
\langle \psi_- ,t|\hat q| \psi_+ ,t\rangle=\bq(t) \quad , \quad 
\langle \psi_- ,t|\hat p| \psi_+ ,t\rangle=\bp(t)
\ee
for any $t$, and the boundary conditions
\be\label{asymptotic vacuum}
\mathop{\lim}_{t\rightarrow\mp\infty}| \psi_\pm ,t\rangle=|0\rangle\, .
\ee

A sketch of the proof of this statement is given in Appendix B. A special role
is played by time-independent $\bp$ and $\bq$, because the previous proposition 
reduces to $\Gamma[\bp,\bq]=-{\cal E}(\bp,\bq)\int\!\!dt$ where ${\cal E}$ is the
usual energy density functional defined by the minimum
\be\label{energy density}
{\cal E}\left(\bp,\bq\right) =
\mathop{\min}_{|\psi\rangle}\, \langle \psi|\hat H| \psi\rangle
\ee
with respect to variations of the time-independent state $|\psi\rangle$ preserving the
time-independent version of the constraints in~\eqref{constraints psi}.

This clearly provides an energy interpretation for the effective Hamiltonian action.
In particular if one evaluates this action on the constant ($\bp$,$\bq$)-values which make 
it stationary, the resulting number is just minus the ``time volume" 
times the ground state energy.
In principle it is possible to compute all the energy levels by means of $\Gamma$, but
higher levels require more work. One possible way is through the two point functions.
In a Hamiltonian framework the propagator splits in the entries of the matrix:
\be\label{propagatorW}
i\langle {\cal T}
\begin{pmatrix}
(q\!-\!\bq)_{t'}(q\!-\!\bq)_t&(p\!-\!\bp)_{t'}(q\!-\!\bq)_t\\
(q\!-\!\bq)_{t'}(p\!-\!\bp)_t&(p\!-\!\bp)_{t'}(p\!-\!\bp)_t\\
\end{pmatrix}
\rangle = W^{(2)}_{t t'}[I,J]=
\begin{pmatrix}
\frac{\delta^2W}{\delta J_{t'}\delta J_t}&\frac{\delta^2W}{\delta I_{t'}\delta J_t}\\
\frac{\delta^2W}{\delta J_{t'}\delta I_t}&\frac{\delta^2W}{\delta I_{t'}\delta I_t}\\
\end{pmatrix}=
\begin{pmatrix}
\frac{\delta \bq_t}{\delta J_{t'}}&\frac{\delta \bq_t}{\delta I_{t'}}\\
\frac{\delta\bp_t}{\delta J_{t'}}&\frac{\delta\bp_t}{\delta I_{t'}}\\
\end{pmatrix}
\ee
(where ${\cal T}$ is the time ordering operator)
so that one could try to think about $p$ and $q$ as different ``fields'' but should also remember 
about the existence of an unusual mixed propagator connecting $p$-legs to $q$-legs or vice versa.
Thanks to~\eqref{eqIJ} one can write this matrix in terms of $\Gamma$ as follows
\be\label{propagatorGamma}
W^{(2)}_{t t'}[I,J]=
\begin{pmatrix}
\frac{\delta \bq}{\delta J}&\frac{\delta \bq}{\delta I}\\
\frac{\delta\bp}{\delta J}&\frac{\delta\bp}{\delta I}\\
\end{pmatrix}_{t t'}=
\begin{pmatrix}
\frac{\delta J}{\delta \bq}&\frac{\delta J}{\delta \bp}\\
\frac{\delta I}{\delta \bq}&\frac{\delta I}{\delta \bp}\\
\end{pmatrix}^{-1}_{t t'}=-
\begin{pmatrix}
\frac{\de^2 \Gamma}{\de \bq\de\bq}&\frac{\de^2 \Gamma}{\de \bp\de\bq}\\
\frac{\de^2 \Gamma}{\de \bq\de\bp}&\frac{\de^2 \Gamma}{\de \bp\de\bp}\\
\end{pmatrix}^{-1}_{t t'}=-\left(\Gamma^{(2)}[\bp,\bq]\right)^{-1}_{t t'}\,\,.
\ee
In order to make the last expression for the two point function more explicit one needs to
invert a matrix whose elements are operators.
In the particular case in which all block entries
of the original matrix are nonsingular, its inverse is given by
\be\label{blockinversion}
\begin{pmatrix}
A&B\\
C&D\\
\end{pmatrix}^{-1}=
\begin{pmatrix}
(A\!-\!B D^{-1}C)^{-1}&(C- D B^{-1}A)^{-1} \\
(B-A C^{-1}D)^{-1}&(D\!-\!C A^{-1}B)^{-1}\\
\end{pmatrix}.
\ee
In our case the operator $W^{(2)}_k$ is symmetric and one
can use the formula in eqn.~\eqref{blockinversion} setting $C=B^T$. 
Let us stress that in order to put the off-diagonal blocks of this inverse 
in the form of eqn.~\eqref{blockinversion} with $C=B^T$ it is only necessary to assume that
$B$ be nonsingular, condition which is met by $\frac{\de^2 \Gamma}{\de \bp\de\bq}$ unless 
$\Gamma$ is extremely pathological.
Once we know how to compute the two point functions by means of $\Gamma$,
we could have access to all the energy gaps $\Delta E_n=E_n-E_0$
through the K\"allen-Lehmann representation of the propagators
\ba\label{KallenLehmann}
\frac{\delta^2W}{\delta I(\tau)\delta I(0)}&=&
\sum_{n\neq 0}e^{-i \Delta E_n\tau}|\langle0|\hat p |n\rangle|^2=
-\sum_{n\neq 0}\int \frac{d E}{2\pi}
e^{-i E\tau}\frac{2 \Delta E_n |\langle0|\hat p |n\rangle|^2}{E^2-\Delta E_n^2+i\epsilon}
\nonumber\\
\frac{\delta^2W}{\delta J(\tau)\delta J(0)}&=&
\sum_{n\neq 0}e^{-i \Delta E_n\tau}|\langle0|\hat q |n\rangle|^2=
-\sum_{n\neq 0}\int \frac{d E}{2\pi}
e^{-i E\tau}\frac{2 \Delta E_n |\langle0|\hat q |n\rangle|^2}{E^2-\Delta E_n^2+i\epsilon}\,.
\nonumber
\ea
Similar expressions hold for mixed derivatives of $W$. 
This tells us that, in principle, by studying the pole structure of the Fourier transformed
two point functions we could compute all the $\Delta E_n$.
As eq.~\eqref{propagatorGamma} shows, this requires the knowledge of the exact
$\Gamma^{(2)}$. In most cases this is not available, and only approximations are possible.
In certain contexts one popular approximation scheme for the computation of the effective
action is the derivative expansion. The zeroth order of such an expansion in the present
Hamiltonian framework can be called the local Hamiltonian approximation (LHA) and
consists of the ansatz: 
$\Gamma=\int\! dt\, \left(\bp\dt\bq-H_{\rm eff}(\bp,\bq)\right)$
where the effective Hamiltonian $H_{\rm eff}$, which is an ultralocal function of its arguments
(i.e. it does not depend on their derivatives), can be computed by setting the fields 
$\bp$ and $\bq$ to constant values.
For this choice, since the second derivatives of $\Gamma$ on constant field configurations commute with each other, 
the inversion rule~\eqref{blockinversion} leads to a simple expression
\ba
\frac{\delta^2W}{\delta I(\tau)\delta I(0)}\!\!\!&=&\!\!\!
-\left[\frac{\de^2 \Gamma}{\de \bp\de\bp}
-\frac{\de^2 \Gamma}{\de \bq\de\bp}\left(\frac{\de^2 \Gamma}{\de \bq\de\bq}\right)^{-1}
\frac{\de^2 \Gamma}{\de \bp\de\bq}\right]^{-1}_{0\tau}
\mathop{=}^{LHA}-\int \frac{d E}{2\pi}
e^{-i E\tau}\frac{\frac{\partial^2 H_{\rm eff}}{\partial \bq\partial\bq}}{E^2-{\rm det}H_{\rm eff}^{(2)}+i\epsilon}
\nonumber\\
\frac{\delta^2W}{\delta J(\tau)\delta J(0)}\!\!\!&=&\!\!\!
-\left[\frac{\de^2 \Gamma}{\de \bq\de\bq}
-\frac{\de^2 \Gamma}{\de \bp\de\bq}\left(\frac{\de^2 \Gamma}{\de \bp\de\bp}\right)^{-1}
\frac{\de^2 \Gamma}{\de \bq\de\bp}\right]^{-1}_{0\tau}
\mathop{=}^{LHA}-\int \frac{d E}{2\pi}
e^{-i E\tau}\frac{\frac{\partial^2 H_{\rm eff}}{\partial \bp\partial\bp}}{E^2-{\rm det}H_{\rm eff}^{(2)}+i\epsilon}
\nonumber\\
\ea
and similar formulae hold for mixed derivatives of $W$. Here 
${\rm det} H^{(2)}=\partial^2_{\bq\bq}H\,
\partial^2_{\bp\bp}H-(\partial^2_{\bq\bp}H)^2$ is the determinant of the Hessian matrix of $H$.
Therefore we see that in the LHA, whenever the second derivatives of $H_{\rm eff}$ commute (as in the
case they are single numbers and not matrices), only one pole appears in the propagators
at the value $({\rm det}H_{\rm eff}^{(2)})^{1/2}$.
Since we are performing a derivative (low energy) expansion, 
in general this pole is the one closer to $E=0$, that is to say the first gap $\Delta E_1$,
unless the matrix elements $\langle0|\hat p |1\rangle$ and $\langle0|\hat q |1\rangle$ vanish.
Therefore we shall use in the LHA approximation the relations
\be
E_0=H_{\rm eff} |_{\rm min} \quad, \quad \Delta E_1=\sqrt{ {\rm det}H_{\rm eff}^{(2)} } \, |_{\rm min} \,.
\ee

So far we have discussed how many properties of a quantum system 
can be deduced from the effective Hamiltonian action,
but how can we compute this action? One way is to use perturbation theory. 
First of all one needs to define propagators and vertex functions.
We already know that in a Hamiltonian framework the propagators of a theory with
Hamiltonian action $\Gamma$ are given by eq.~\eqref{propagatorGamma}.
The vertex functions generated by $\Gamma$ are simply given by:
\be\label{vertex}
\frac{\delta^m}{\delta \bp^m}\frac{\delta^n \Gamma}{\delta \bq^n}\bigg|_{\bq=\bp=0}\, ,\quad m+n>2
\ee
and therefore generically comprehend $m$ $p$-legs and $n$ $q$-legs.
Since perturbation theory in phase space is built on tree level propagators and vertices, 
one can read off these ingredients from \eqref{propagatorGamma} and \eqref{vertex}
by substituting $\Gamma$ with the bare action $S$. 
For instance, to get the one-loop result one changes variables of integration in~\eqref{intdiffeq}
according to $p=\bp+\hbar^\half p^\prime$, $q=\bq+\hbar^\half q^\prime$,
and Taylor-expands both $S$ and $\Gamma$ around $\hbar=0$ up to linear terms
\ba
S(p,q)\!\!&=&\!\!S(\bp,\bq)+
\frac{\hbar}{2} (p^\prime,q^\prime) S^{(2)}(\bp,\bq)(p^\prime,q^\prime)^T 
+ o(\hbar^2)\nonumber\\
\Gamma[\bp,\bq]\!\!&=&\!\!\Gamma_0[\bp,\bq]+\hbar \Gamma_1[\bp,\bq]+ o(\hbar^2)\nonumber \, .
\ea
The change of variable goes along with a change of measure, due to the Jacobian determinant ${\rm Det}\hbar$, 
such that the new measure becomes $\mu[p^\prime,q^\prime]={\rm Det}\frac{1}{2\pi}$. 
The Gaussian path integral over $p^\prime$ and $q^\prime$ combined with such a measure gives 
$\Gamma_1[\bp,\bq]=\frac{i}{2}\log{\rm Det} \left(-i S^{(2)}[\bp,\bq]\right)$, where $S$ is the bare Hamiltonian action
(together with the obvious result $\Gamma_0[\bp,\bq]=S[\bp,\bq]$).
The block determinant can be written in a more explicit form by means of
the general formula
\be\label{blockdet}
{\rm det}\begin{pmatrix}
A&B\\
C&D\\
\end{pmatrix}=
{\rm det}A\,\,{\rm det}(D-C A^{-1} B)={\rm det}D\,\,{\rm det}(A-B D^{-1} C)
\ee
where the first expression is true if ${\rm det}A\neq 0$ and the second if ${\rm det}D\neq 0$.
Therefore, if $\frac{\de^2 S}{\de \bp \de \bp}$ is non-vanishing
\ba\label{1loop}
\Gamma_1[\bp,\bq] \!\!\!\!&=&\!\!\!\! \frac{i}{2}\log{\rm Det}\!\!\left[-\frac{\de^2 S}{\de \bp \de \bp} \left( \frac{\de^2 S}{\de\bq \de\bq}
-\frac{\de^2 S}{\de\bp \de\bq}\!\!\left(\frac{\de^2 S}{\de\bp \de\bp}\right)^{-1}\!\!\!
\frac{\de^2 S}{\de\bq \de\bp}\right) \right]\nonumber\\
&=&\!\!\!\!\frac{i}{2}\log{\rm Det}\!\!\left[\left(\!-\dt^2\!-{\rm det} H^{(2)}+
\left(\dt\frac{\pa^2H}{\pa \bp \pa \bq}\right)
\!+\! \left(\!\dt\log \frac{\pa^2H}{\pa \bp \pa \bp}\right)\!\!
\left(\!\dt-\frac{\pa^2H}{\pa \bp \pa \bq}\right)\right)\!\de\right]\,\,\,
\ea
which reduces to the usual one-loop formula for the effective action in the case of
a bare Hamiltonian like the one in \eqref{QMbarehamiltonian}. In the formula above we have used the symbol $\delta$ for $\delta(t-t')$ which defines the operator.
If instead $\frac{\de^2 S}{\de \bp \de \bp}$ vanishes while
$\frac{\de^2 S}{\de \bq \de \bq}$ is non-vanishing, the result can be obtained 
from \eqref{1loop} by replacing $\de_\bp$ with $\de_\bq$ and vice versa.

In the rest of this paper we will work on a non-perturbative setting for the computation of the effective Hamiltonian action 
and we will choose $\hbar$ as our unit of action.

\subsection{The average effective Hamiltonian action}
A non-perturbative definition of the path integral~\eqref{intdiffeq}
can be given by a functional RG flow equation.
The starting point of this construction is the introduction of an external parameter in the theory. 
This allows to reduce the task of computing  the functional integral in the simpler task of computing its infinitesimal variation under changes of such a parameter.
In quantum mechanics the external parameter can be dimensionless, since the number of degrees of freedom is finite and no regularization is needed.
Instead the generalization of the construction to field theories requires the introduction of
a dimensionful parameter $k$, such that its variation corresponds to a {\it coarse graining} operation (otherwise we meet infinities in the computation of the infinitesimal variation). 
An alternative way is to assume that the theory has already been regularized, as for example by the introduction of a UV cutoff $\Lambda$, in which case it is possible to deal with a dimensionless parameter also in field theories 
(related to the ratio between the dimensionful $k$ and $\Lambda$).
Since by varying $k$ we will get a one parameter flow of theories, we will need initial conditions in order to integrate it. A convenient way to deal with this issue is to choose the dependence on $k$ in such a way that the flow interpolates between full functional integration (conventionally at $k=0$) and no integration at all (conventionally at $k=\Lambda$, even if $\Lambda$ might in some cases be displaced at $+\infty$).
The no integration limit can also be realized considering $k$ as a mathematical parameter unrelated to a physical sounding coarse-graining procedure,
and, in the presence of the physical UV cutoff $\Lambda$, taking the limit $k\to\infty$.
Sticking to this framework we introduce such a parameter, by means of a modification of the bare action and of the functional measure
\be\label{hampathint2}
e^{i W_k[I,J]}=\int \left[ dp dq\right]\mu_k[p,q] 
e^{\, i\left\{S[p,q]+\Delta S_k[p,q] + I\cdot p + J\cdot q  \right\} }
\ee
and ask for $\mu_k \exp\{i \Delta S_k\}$ to become $\mu$ as $k\rightarrow 0$ and
to provide a rising delta functional as $k\rightarrow \Lambda$.
As traditional, to keep the framework as simple as possible, we choose $\Delta S_k$
to be quadratic in the fields
\be\label{deltaSk}
\Delta S_k[p,q]=\half (p,q)\cdot R_k \cdot(p,q)^T 
\ee
such that we need $R_k\rightarrow 0$ and $\mu_k\rightarrow \mu$ when $k\rightarrow 0$, as well as $R_k\rightarrow \infty$ and $\mu_k\rightarrow \left({\rm Det} \frac{R_k}{2\pi}\right)^\half$ when $k\rightarrow \Lambda$.
These constraints can be satisfied by several choices for the symmetric matrix $R_k$ and for the measure $\mu_k$.
In this paper we will consider only two simple cases in which the only non-vanishing
entries of $R_k$ are either off-diagonal and built out of an odd differential operator or diagonal and built out of even differential operators. These respectively read
\ba\label{offdiagR}
R_k(t,t')&=&
\begin{pmatrix}
0 & r_k(-\dt^2)\dt \de(t-t') \\
-r_k(-\dt^2)\dt \de(t-t') & 0\\
\end{pmatrix}
\\
\label{diagR}
R_k(t,t')&=&
\begin{pmatrix}
{\cal R}_k^p(-\dt^2)\de(t-t') & 0 \\
0 & {\cal R}_k^q(-\dt^2)\de(t-t')\\
\end{pmatrix}
\ea
The first choice can be interpreted as a $k$-dependent deformation of the
symplectic potential $\lambda=p dq$, by means of an operator
$(1+r_k)$ which, after the pull-back by a section defining the specific path, might become a differential operator. This interpretation suggests the appropriate $k$-dependent
deformation of the functional measure: if the new symplectic potential is 
$\lambda_k=p(1+r_k) dq$,
the new non-trivial Liouville measure would become
$\mu_k= \left({\rm Det}\frac{\sigma_k}{2\pi}\right)^\half$, where $\sigma_k=d\lambda_k$ is the regularized symplectic form.
This choice for the measure indeed provides the correct normalization of the  Gaussian
rising delta functional~\cite{VZ}.
Following this line of thought we can guess a convenient choice for the regularized measure
also in the second case of a diagonal regulator. 
The straightforward adaptation of the previous argument is insisting in adding to the 
fundamental symplectic matrix our regulator matrix, and then taking its determinant.
To summarize, the regularized functional measures we will use together with the
regulators \eqref{offdiagR} and \eqref{diagR} respectively are
\ba\label{offdiagmu}
\mu_k&=&\left[{\rm Det}\frac{1}{2\pi}
\begin{pmatrix}
0 & \left(1+r_k(-\dt^2)\right)\dt \de(t-t') \\
-\left(1+r_k(-\dt^2)\right)\dt \de(t-t') & 0\\
\end{pmatrix}\right]^\half
\\
\label{diagmu}
\mu_k&=&\left[{\rm Det}\frac{1}{2\pi}
\begin{pmatrix}
{\cal R}_k^p(-\dt^2)\de(t-t') & \dt \de(t-t') \\
-\dt \de(t-t') & {\cal R}_k^q(-\dt^2)\de(t-t')\\
\end{pmatrix}\right]^\half \,.
\ea

The definition of the average effective Hamiltonian action (AEHA)
$\Gamma_k[\bp,\bq]$ is
\be
\Gamma_k\left[\bp,\bq\right] +\Delta S_k\left[\bp,\bq\right]=
\mathop{\rm ext}_{I,J}\left( W_k[I,J] -I \cdot\bp -J\cdot \bq\right)\nonumber\,.
\ee
Note that the sources minimizing the r.h.s. will in general depend on $k$.
Again it is easy to write an integro-differential equation for the AEHA:
\be\label{intdiffeqk}
e^{i \Gamma_k[\bp,\bq]}=\int \left[ dp dq\right]\mu_k[p,q]  
e^{\, i\left\{S[p,q]+\Delta S_k[p-\bp,q-\bq]
-(p-\bp)\frac{\de \Gamma_k}{\de \bp}-(q-\bq)\frac{\de \Gamma_k}{\de \bq}\right\} }\,.\\
\ee
When $k\rightarrow 0$ eq.~\eqref{intdiffeqk} trivially reduces to
eq.~\eqref{intdiffeq} and the AEHA becomes the full effective Hamiltonian action.
It is not hard to check that when $k\rightarrow \Lambda$ the r.h.s. of
eq.~\eqref{intdiffeqk} reduces to $\exp\{i S[\bp,\bq]\}$ and the AEHA coincides with the bare Hamiltonian action.
A sketch of the proof can be found in Appendix C.

The relation between the average effective Hamiltonian and Lagrangian actions
is the same as for the full effective actions:
\be\label{AELA}
\Gamma_k[\bq]=\mathop{\rm ext}_{\bp}\Gamma_k[\bp,\bq] \, .
\ee

We observe that this is evident in the simplest possible case, i.e. when the bare action is quadratic in the momenta, as in~\eqref{QMbarehamiltonian},
since $\frac{\pa^2 H}{\pa p^2} $ and $\frac{\pa^2 H}{\pa p \pa q}$ are constant (the latter is actually zero).
Indeed the integration over $p$ in~\eqref{intdiffeqk} can be performed exactly and in such a case 
one discovers that also the AEHA must be quadratic in the momenta and that
for any $k$ the canonical momentum that extremizes it is $\bp=\dt\bq$.
As a result, plugging this field configuration in~\eqref{intdiffeqk},
using the definition~\eqref{AELA} and integrating out the momenta, one obtains
\be\label{Lintdiffeqk}
e^{i \Gamma_k[\bq]}=\int \left[dq\right]\mu_k[q]  
e^{\, i\left\{S[q]+\Delta S_k[q-\bq]-(q-\bq)\frac{\de \Gamma_k}{\de \bq}\right\} }\\
\ee
where now $\mu_k[q]\equiv\int\left[dp\right]\mu_k[p,q]e^{\,-i\frac{p^2}{2} }$
and $\Delta S_k[q]$ arises from the chosen $\Delta S_k[p,q]$.
For example, if one adopts the scheme of eqs.~\eqref{offdiagR} and~\eqref{offdiagmu}
then 
\ba
\mu_k[q]&=&\left[{\rm Det}\frac{1}{2\pi}\left(1+r_k(-\dt^2)\right)^2(-\dt^2)\de\right]^\half
\nonumber\\
\Delta S_k[q]&=&\half \dt q\cdot(r_k^2+2r_k) \dt q\nonumber\, .
\ea
As usual, the $k\rightarrow\Lambda$ limit of the average effective Lagrangian action
coincides with the bare Lagrangian action while the $k\rightarrow\Lambda$ limit
gives the full quantum effective Lagrangian action.

In this work we are interested in the cases which depart from such a simple situation.

\subsection{RG flow equation for the AEHA}

In this section we discuss the translation of the functional 
integro-differential equation~\eqref{intdiffeqk}
in a functional differential equation describing a flow parameterized by $k$.

Denoting by ``.'' the operation $k\partial_k$, and
acting with it on eq.~\eqref{intdiffeqk} one obtains
\be\label{preflow1}
i\dot{\Gamma}_k=\frac{\dot{\mu}_k}{\mu_k} +i\langle
\dot{\Delta S}_k[p-\bp,q-\bq]\rangle_k\,.\nonumber
\ee
Since $\Delta S_k$ has been chosen quadratic in the fields, the expectation
value can be rewritten by means of the
$k$-dependent version of formulae~(\ref{propagatorW},\ref{propagatorGamma}).
Denoting $\tilde{\Gamma}_k\left[\bp,\bq\right]\equiv\Gamma_k\left[\bp,\bq\right] +\Delta S_k\left[\bp,\bq\right]$, these read
\ba
i\langle {\cal T}\!
\begin{pmatrix}
(q\!-\!\bq)_{t'}(q\!-\!\bq)_t&(p\!-\!\bp)_{t'}(q\!-\!\bq)_t\\
(q\!-\!\bq)_{t'}(p\!-\!\bp)_t&(p\!-\!\bp)_{t'}(p\!-\!\bp)_t\\
\end{pmatrix}
\rangle_k 
= W^{(2)}_{\!k\phantom{(2}t t'}[I,J]=\!
\begin{pmatrix}
\frac{\delta^2W_k}{\delta J_{t'}\delta J_t}&\frac{\delta^2W_k}{\delta I_{t'}\delta J_t}\\
\frac{\delta^2W_k}{\delta J_{t'}\delta I_t}&\frac{\delta^2W_k}{\delta I_{t'}\delta I_t}\\
\end{pmatrix}=\nonumber\\
\!=\! \begin{pmatrix}
\frac{\delta \bq_t}{\delta J_{t'}}&\frac{\delta \bq_t}{\delta I_{t'}}\\
\frac{\delta\bp_t}{\delta J_{t'}}&\frac{\delta\bp_t}{\delta I_{t'}}\\
\end{pmatrix}
=\begin{pmatrix}
\frac{\delta J}{\delta \bq}&\frac{\delta J}{\delta \bp}\\
\frac{\delta I}{\delta \bq}&\frac{\delta I}{\delta \bp}\\
\end{pmatrix}^{-1}_{t t'}=-
\begin{pmatrix}
\frac{\de^2 \tilde{\Gamma}_k}{\de \bq\de\bq}&\frac{\de^2 \tilde{\Gamma}_k}{\de \bp\de\bq}\\
\frac{\de^2 \tilde{\Gamma}_k}{\de \bq\de\bp}&\frac{\de^2 \tilde{\Gamma}_k}{\de \bp\de\bp}\\
\end{pmatrix}^{-1}_{t t'}
=- \left(\tilde{\Gamma}^{(2)}_k[\bp,\bq]\right)_{t t'}^{-1}\,.
\nonumber
\ea
Therefore, for any quadratic regulator, eq.~\eqref{preflow1}
can be written as
\be\label{preflow2}
i\dot{\Gamma}_k=\frac{\dot{\mu}_k}{\mu_k} -
\half {\rm Tr}\left[\left(\Gamma_k^{(2)}+R_k\de\right)^{-1}\dot{R}_k\de\right]
\ee
where $R_k\de=\Delta S_k^{(2)}$.
Here one has still freedom for the choice of $\mu_k$ as a functional of $R_k$.
By using the inversion formula~\eqref{blockinversion}
one can find a more explicit form for the flow equation.
Adopting the regulator~\eqref{offdiagR}
affecting only the Legendre transform term of the bare action 
(i.e. the symplectic potential) and the corresponding minimally 
deformed Liouville functional measure~\eqref{offdiagmu},
 eq.~\ref{preflow2} becomes 
\ba\label{offdiagflow}
i\dot{\Gamma}_k&=&{\rm Tr}\left[\dot{r}_k\left(1+r_k\right)^{-1}\de\right]\nonumber\\
&-&{\rm Tr}\left[(\dot{r}_k\partial\de)\left(
\left(r_k\partial\de+\frac{\de^2\Gamma_k}{\de \bq\de\bp}\right)
-\frac{\de^2\Gamma_k}{\de \bp\de\bp}
\left(-r_k\partial\de+\frac{\de^2\Gamma_k}{\de \bp\de\bq}\right)^{-1}
\frac{\de^2\Gamma_k}{\de \bq\de\bq}
\right)^{-1}
\right]\,\,.
\ea
where we denote $(\partial\de)_{t_1 t_2}=\partial_{t_1}\de(t_1-t_2)$.
Instead, the choice of a diagonal regulator~\eqref{diagR} and of the corresponding
measure~\eqref{diagmu} leads to the flow equation
\ba\label{diagflow}
i\dot{\Gamma}_k&=&
\half{\rm Tr}\left[(\dot{\cal R}_k^p\de)\left(
{\cal R}_k^p\de -\left(\partial\de\right)
\left({\cal R}_k^q\de\right)^{-1}
\left(-\partial\de\right)
\right)^{-1}
\right]
\nonumber\\
&+&\half{\rm Tr}\left[(\dot{\cal R}_k^q\de)\left(
{\cal R}_k^q\de-\left(-\partial\de\right)
\left({\cal R}_k^p\de\right)^{-1}
\left(\partial\de\right)
\right)^{-1}
\right]\nonumber\\
&-&\half{\rm Tr}\left[(\dot{\cal R}_k^p\de)\left(
\left({\cal R}_k^p\de+\frac{\de^2\Gamma_k}{\de \bp\de\bp}\right)
-\frac{\de^2\Gamma_k}{\de \bq\de\bp}
\left({\cal R}_k^q\de+\frac{\de^2\Gamma_k}{\de \bq\de\bq}\right)^{-1}
\frac{\de^2\Gamma_k}{\de \bp\de\bq}
\right)^{-1}
\right]\nonumber\\
&-&\half{\rm Tr}\left[(\dot{\cal R}_k^q\de)\left(
\left({\cal R}_k^q\de+\frac{\de^2\Gamma_k}{\de \bq\de\bq}\right)
-\frac{\de^2\Gamma_k}{\de \bp\de\bq}
\left({\cal R}_k^p\de+\frac{\de^2\Gamma_k}{\de \bp\de\bp}\right)^{-1}
\frac{\de^2\Gamma_k}{\de \bq\de\bp}
\right)^{-1}
\right]
\,\,.
\ea
Notice that, thanks to the regularization of the functional measure, these equations
correctly reproduces the non-renormalization of $H_k$ in the trivial cases in which the bare 
Hamiltonian either vanishes or depends on just one field out of $p$ and $q$.
As far as the reality properties of this equation are concerned, there is no difference
with the standard Lagrangian formalism in real time, that is to say, the imaginary unit on the l.h.s.
is needed in order to ensure reality of $\Gamma_k$. This is because in real time the traces on the r.h.s.
usually are integrals of functions with poles on the real axis, which thus lead to imaginary values. An appropriate prescription
should be given in order to displace these poles off the real axis.
As usual in QFT one adopts the prescription which relates the Minkowskian theory to the Euclidean theory by a continuous Wick rotation.
The same can be done in QM. 
The reader can find more details about this translation to imaginary time in appendix D.

The previous flow equations are still too general for a first approach to their meaning and
application, therefore let us give more specific and simple forms of the first one of them,
eq.~\eqref{offdiagflow}.
As a first example let's consider the
truncation $\Gamma_k=\int\! dt\, \left(\bp\dt\bq-\frac{1}{2} \bp^2-V_k(\bq)\right)$.
Introducing the notation $P_k(-\dt^2)=-\dt^2 (1+r_k)^2$
one finds the RG flow equation
\be
-i\int\!\! dt\,\dot{V}_k(\bq)=\frac{1}{2}{\rm Tr}\left[\dot{P}_k P_k^{-1}\right]
-\frac{1}{2}{\rm Tr}\left[\dot{P}_k\left(P_k-V_k^{(2)}(\bq)\right)^{-1}\right]
\ee
which is what one gets by the effective average Lagrangian action approach~\cite{VZ} 
in the local potential approximation (LPA).
A more general example is the local Hamiltonian approximation (LHA), i.e.
the case in which the flow equation for the truncation 
$\Gamma_k=\int\! dt\, \left(\bp\dt\bq-H_k(\bp,\bq)\right)$
is evaluated on constant $\bq$ and $\bp$ configurations.
For this choice, if the second derivatives of $\Gamma_k$ commute with each other as 
in the present case where they are 1-by-1 bosonic matrices, the operators 
in the trace can be simplified and one obtains
\ba\label{lha_flow_r}
-i\int\!\! dt\,\dot{H_k}(\bp,\bq)=
&-&{\rm Tr}\left[\left(\frac{\dot{r}_k}{1+r_k}\de\right) \frac{{\rm det} H_k^{(2)}(\bp,\bq)}{-\partial^2 (1+r_k)^2\de-{\rm det} H_k^{(2)}(\bp,\bq)}\right]\nonumber\\
&+&{\rm Tr}\left[\frac{\left(\dot{r}_k\partial\de\right) \frac{\de^2 H_k}{\de\bp\de\bq}(\bp,\bq)}{-\partial^2 (1+r_k)^2\de-{\rm det} H_k^{(2)}(\bp,\bq)}\right]
\ea
where 
${\rm det} H_k^{(2)}=\partial^2_{\bq\bq}H_k\,
\partial^2_{\bp\bp}H_k-(\partial^2_{\bq\bp}H_k)^2$ is the determinant of the Hessian matrix of $H_k$.
Notice that the second trace vanishes whenever it is possible to evaluate it in Fourier space
and when the domain in such space is symmetric around the origin. If this is the case we are left with
\be\label{lha_flow_P}
i\int\!\! dt\,\dot{H_k}(\bp,\bq)=
\half{\rm Tr}\left[\frac{\dot{P}_k}{P_k}\de\,\frac{{\rm det} H_k^{(2)}(\bp,\bq)}
{P_k\de-{\rm det} H_k^{(2)}(\bp,\bq)}\right]\,.
\ee
Here one could adopt any of the regulators ${\cal R}_k$ developed in the vast literature about the average effective Lagrangian action~\cite{BTW,litim},
and plug it in the last formula by $P_k(-\dt^2)=-\dt^2+{\cal R}_k(-\dt^2)\,$.
One of the simplest choices for the regulator is a constant $r_k$, that is to say
an operator which is multiplicative in both time and frequency representations;
in other words a function of $k$ and $\Lambda$ only.
If no UV cutoff is present, this choice is possible only in quantum mechanics,
because it does not produce any coarse graining and therefore it does not regularize
the functional traces.
Assuming $\dot{r}_k>0,\, \forall k \in (0,\Lambda),$ one can trade $k$
for the dimensionless parameter $r_k$. 
Thus, in LHA and if the second derivatives of $H_k$ commute with each other,
assuming that the traces can be written as $\int\!\! dt\int\!\!\frac{dE}{2\pi}$ (after Fourier transform),
and that there is no UV cutoff in the theory, 
then by Wick rotating the trace ($E\to iE$) one gets
\be\label{lha_flow_r}
\frac{d H_r}{d r}=
-\frac{1}{2 (1+r)^2} \left( {\rm det} H_r^{(2)} \right)^\half\,.
\ee
A different choice which makes the computation of the traces even simpler than for 
a constant $r_k$ is the square root of the Litim regulator~\cite{litim}.
Denoting by $r_k(E^2)E$ the Fourier transform of $r_k(-\dt^2)i\dt$, and with
$\theta$ the Heaviside step function, after Wick rotation such a regulator reads
\be\label{sqlitim}
r_k(E^2)E=-(k+E)\theta(k+E)\theta(-E)+(k-E)\theta(k-E)\theta(E)\, .\nonumber
\ee
In the LHA and if the second derivatives of $H_k$ commute with each other,
this gives the same result as~\eqref{lha_flow_P} for $P_k(E^2)=k^2\theta(k^2-E^2)+E^2\theta(E^2-k^2)\,$,
that is
\be\label{lha_flow_opt}
\dot{H}_k=
-\frac{k}{\pi}\frac{{\rm det} H_k^{(2)}}{k^2+ {\rm det} H_k^{(2)}}\,.
\ee
Of course if one considers $H_k(\bp,\bq)=T_k(\bp)+V_k(\bq)$ as an initial condition for the flow, whenever both $T_k$ and $V_k$ are polynomials of degree higher than two,
the determinant becomes a function of both $\bq$ and $\bp$ so that the flow generates also mixed $\bp$ and $\bq$ dependence in the effective Hamiltonian. 
Therefore one should consider a larger truncation in order to track such terms. 
Also a structure of a $\sigma$-model kind, quadratic in the momenta,
generates a dependence in the momenta which is more than quadratic.
We stress that in general the flow will generate also a dependence
 on time derivatives of $q$ and $p$ variables.
This goes beyond the LHA but it is still compatible with the standard Hamiltonian
approach as long as one starts the flow at the UV with a derivatives-free bare Hamiltonian.

\subsection{Exercise: the ground state energy and gap of models that are 
more than quadratic in the momenta}
As an example of the application of the framework discussed in the previous subsections to specific problems,
we will present the computation of the first two energy levels of few exactly solvable systems for which no simple
Lagrangian description is available, due to to the fact that the functional integral over the conjugate momenta is not Gaussian.
This will serve as a check of the soundness of the formalism, but the reader is invited to remember that the very same simple 
computations explained in the following would work also for much more complicated models. Let us recall that the functional RG has already
 been successfully applied to the computation of the spectrum of quantum
mechanical models in the configuration space formulation~\cite{AHTT,gies}.

The systems we are going to address have the following classical 	 Hamiltonian:
\be
H_n(p,q)=\left(\frac{p^2+\omega^2 q^2}{2}\right)^n\,.
\label{powerHO}
\ee
They are easy to solve due to the $O(2)$ symmetry which forces the Hamiltonian to depend
only on the ``action" and not on the ``angle" coordinate in phase space.
Even without performing a canonical transformation to such coordinates, the
energy spectrum can be built by ladder operators. Rescaling the variables
$q=q'/\sqrt{\omega}$ and $p=\sqrt{\omega}p'$
as well as the Hamiltonian $H=\omega^n H'$ we can reduce the problem to
the one with $\omega=1$, therefore in the following we will restrict to 
such a case. The operator algebra of these quantum
models is completely described by
\be
\hat a=\frac{\hat q+i\hat p}{\sqrt{2}} \quad , \quad 
\hat a^\dag=\frac{\hat q-i\hat p}{\sqrt{2}}\quad , \quad
\hat a \hat a^\dag - \hat a^\dag \hat a=1\, .
\ee
The Hamiltonian operator is just the $n$-th power of $\left(\hat N+\half\right)$
where $\hat N = \hat a^\dag \hat a$ is the number operator. This is enough
to deduce the whole energy spectrum for any positive integer $n$.

In order to reproduce such a spectrum by means of the RG flow equation, the first step
is to specify the initial condition for the integration of the flow.
From the discussion of the previous subsections follows that the most
suitable initial condition is $\Gamma_{k=\Lambda}=S$, where $S$
is the bare action to be inserted in a path integral, as the input specifying
which system is being studied.
At this point it is necessary to recall that such a bare action is in
one-to-one correspondence with the Hamiltonian of the operator representation:
the bare Hamiltonian is just the Weyl symbol of the Hamiltonian operator.
Let us remind that an operator $\hat O(\hat p, \hat q)$, 
can always be written as a sum of symmetrized (in $\hat p$ and $\hat q$) operators
\be
\hat O= {\hat O}_S+\sum_i {\hat O}_{i S}=\hat O_W
\label{weyl1}
\ee
which is what one calls the Weyl-ordered version of $\hat O$.
Also, its average on coordinate $\hat q$ eigenstates with eigenvalues $x$ and $y$ is conveniently given by
\be
\langle x | \hat O | y \rangle=\int dp\,  \langle x |p \rangle\,O_W\!\left(p,\frac{x+y}{2}\right) \langle p |y \rangle \,.
\label{weyl2}
\ee
The function $O_W$ in the right hand side of eq.~\eqref{weyl2} is called the Weyl
symbol of $\hat O$, and it can be considered as the classical counterpart of $\hat O$.
There are many ways to compute this function; one is to Weyl-order 
$\hat O$ and then to replace the operators in $\hat O_W$ with c-numbers.
Another way is through the relation
\be
O_W(p,q)=\int\!\! dx\, e^{i p x} \langle q-\frac{x}{2} | \hat O (\hat p,\hat q) | q+\frac{x}{2} \rangle
\ee
where the bra's and ket's are again eigenstates of the $\hat q$ operator.
For instance, considering the models in Eq.~\eqref{powerHO}, in the $n=2$ and $n=3$ cases such symbols read
\be
H_{2W}(p,q)=\left(\frac{p^2+q^2}{2}\right)^2-\frac{1}{4}\quad , \quad
H_{3W}(p,q)=\left(\frac{p^2+q^2}{2}\right)^3-\frac{5}{4}\left(\frac{p^2+q^2}{2}\right) \,.
\ee
Notice that both subtraction terms above, due to Weyl ordering, are proportional
to $\hbar^2$, but in natural units such a dependence disappears.

Inserting these initial conditions in the flow equation for the LHA
one can compute the full quantum effective Hamiltonian at $k=0$.
Such a task can be performed by numerically integrating the flow equation.
However, if one is interested in simple quantities as the first two energy levels,
this might be unnecessary: it could be enough
to truncate the LHA to a polynomial in $z\equiv(p^2+q^2)/2$ of finite order.
Indeed if the bare Hamiltonian depends on $p$ and $q$ only through $z$,
in the LHA approximation also $H_k$ can be shown to respect this symmetry, for suitable cutoff operators.

We started by studying these polynomial truncated flows as generated by
equations~\eqref{lha_flow_r} and \eqref{lha_flow_opt} finding that 
singularities appear at nonvanishing values of $k$.
This happens because at some $k$ the radius of convergence 
of the necessary expansion of the r.h.s. in powers of $z$ goes to zero,
 a fact related to the vanishing of the terms quadratic in the fields 
in the bare Hamiltonian of the $n=2$ model.
If no expansion is performed, as in the numerical integration of the flow equation for $H_k$,
 no singularity is met and the ground state and gap can be
estimated by the value of $H_k$ and of $({\rm det} H_k^{(2)})^{1/2}$ at the minimum.
However these estimates do not reach a great accuracy either because of spurious dependence
on the boundary conditions (which can be controlled by some nonlinear redefinitions of $H_k$)
or because of numerical errors: typically we reached no more than two digit accuracy in the region around the minimum.
In order to get stable predictions with a precision better than $1\%$ we turned to
a different choice of regulators, curing the problem about the polynomial expansion
of the flow equation. Such a choice is that of a diagonal regulator,
as in eq.~\eqref{diagflow}. 
We chose this regulator to be constant, i.e. ${\cal R}_k^p={\cal R}_k^q={\cal R}$ a multiplicative operator (recall that we are assuming
$\omega^2=1$ therefore even if ${\cal R}_k^p$ and ${\cal R}_k^q$
have different dimensions we can set them equal if we assume their ratio
to be some power of $\omega$).
We also introduced a UV cutoff $\Lambda$ in order to
control the convergence of the flow for ${\cal R}\rightarrow\infty$.
As a result we observed that, for such a constant regulator, $\Lambda$ can be
removed only after the integration of the flow from  ${\cal R}=\infty$ to  
${\cal R}=0$.
The resulting flow equation in the LHA is
\be
\partial_{\cal R}\dot{H}_{\cal R}=
-\frac{1}{\pi}\arctan\left(\frac{\Lambda}{{\cal R}}\right)+
\frac{2{\cal R}+\partial^2_{\bp\bp}H_{\cal R}+\partial^2_{\bq\bq}H_{\cal R}}{2\pi{\cal D}_{\cal R}}
\arctan\left(\frac{\Lambda}{{\cal D}_{\cal R}}\right)
\ee
where we defined
\be
{\cal D}_{\cal R}=\sqrt{{\cal R}^2 
+{\cal R}\left(\partial^2_{\bp\bp}H_{\cal R}+\partial^2_{\bq\bq}H_{\cal R}\right)
+{\rm det}H_{\cal R}^{(2)}}\nonumber\,.
\ee
In this scheme good estimates for the ground state energy $E_0$ and the energy gap $\Delta E_1=E_1\!-\!E_0$ can be obtained
by simple polynomial truncations.
For a bare Hamiltonian which is a polynomial of order $n$ we consider two cases: a truncation with a polynomial of the same
order $n$ and another of order $n+1$.
In the latter case we add a suffix ${}^{+1}$ to the corresponding quantities $E_0^{+1}$ and $\Delta E_1^{+1}$. We give the results
obtained by choosing as an initial condition both the Weyl-ordered $H_{nW}$ and the Weyl-uncorrected Hamiltonian $H_{n}$:
\begin{center}
\begin{tabular}{|l|l|l|l|l|l|l|l|l|l|l|l|l|}
 \hline
 Bare Hamiltonian& $E_0^{\rm exact}$ & $E_0$ & $E_0^{+1}$&  $\Delta E_1^{\rm exact}$ & $\Delta E_1$ & $\Delta E_1^{+1}$ \\
 \hline
 $H_{2W}$ & 1/4 & 0.24936 & 0.24936 & 2& 1.99871 & 1.99871 \\ 
$H_2$ & 1/2 & 0.49989&  0.49994 &  2& 1.99867 & 1.99985  \\ 
$H_{3W}$ & 1/8 & 0.12492& 0.124886 & 13/4 & 3.24736& 3.24905 \\
$H_3$ & 3/4 & 0.749849& 0.74856   & 9/2 &4.4991 & 4.4939 \\
 \hline
\end{tabular}
\end{center}
We note that the quantities $E_0$ and $\Delta E^1$ depend on the local properties of the effective
Hamiltonian at the minimum ($\bar p=\bar q=0$) and therefore can be extracted with a good approximation adopting simple polynomial truncations.
From the table we see that there is no clear pattern on the change of the precision of the results when increasing the order of the truncation. 
In the worst case we find a relative error of order $10^{-3}$. In order to achieve a better accuracy, going to next-to-leading order  
in the derivative expansion would probably do the job.

We remark that for the first time in the functional RG approach one faces the ordering problem in the choice of the bare Hamiltonian 
function which corresponds to the initial condition for the flow.
This feature generally extends to QFT, therefore one needs to keep it
in mind before interpreting the results obtained by choosing an
initial condition which is non-separable in $p$ and $q$.

\subsection{The average effective Hamiltonian action in fermionic quantum mechanics}
Since fermions usually have a first order dynamics, 
the Hamiltonian formulation of it is identical to the Lagrangian one.
Therefore the AEHA formalism in this case is identical to the traditional Lagrangian approach.
For completeness we will briefly review it in this subsection.

Let's consider as an example a free system
whose Lagrangian variables are $n$ real Grassmann-valued functions of time:
$\left\{\theta^i(t)\right\}_{i=1,...,n}$\ , evolving according to the following Lagrangian:
\be\label{spinorL}
{\cal L}(\theta(t),\partial_t\theta(t))=\frac{1}{2}\theta^i(t)i\partial_t\theta^j(t)\delta_{ij}-V(\theta^i(t)) \,.
\ee
Defining the momenta $\pi_i$ as the right partial derivatives of $L$ with respect to $\dt\theta^i$
we find $n$ second class primary constraints:
\be\label{constraints}
\chi_i(t) = \pi_i(t)-\frac{i}{2}\delta_{i j}\theta^j(t)=0 
\ee
which cause the canonical Hamiltonian $H=\pi_i\dt\theta^i -L=V(\theta^i)$ to be independent of $\pi_i$.
The relevant phase space is the surface $\cal{S}$ defined by ($\ref{constraints}$), 
a complete set of independent coordinates on it
is given by $\theta^i$ and the functional integral is to be taken over 
all paths $\theta^i(t)$ lying on this surface.
In presence of second class constraints and assuming that the whole phase space
is endowed with a symplectic structure $\sigma=d\lambda$, 
we can define a nondegenerate symplectic form 
$\tilde{\sigma}=\tilde{d}\tilde{\lambda}$ on the reduced phase space,
simply by restricting $\sigma$ to $\cal{S}$. 
As the inverse of $\sigma$ is the Poisson bracket $[\, ,\, ]$, 
the inverse of $\tilde{\sigma}$ is the Dirac bracket $[\, ,\, ]_{\sim}$,
which in the reduced phase space coordinates $\theta^i$ has components:
$[\theta^i,\theta^j]_{\sim}=-i\delta^{ij}=[\chi_i,\chi_j]$ .
The kinetic term in \eqref{spinorL} can be interpreted 
as the Legendre transform term on $\cal{S}$,
i.e. as the pullback of the symplectic potential $\tilde{\lambda}$ by a section $\theta^i(t)$.
The appropriate measure for functional integration over $\cal{S}$ is again 
the square root of the superdeterminant of the symplectic form $\tilde{\sigma}$~\cite{HT}.
In conclusion the functional integral over the reduced phase space reads
\be\label{thetaint1}
 Z = \int [d\theta] \mu[\theta] e^{\,i S[\theta]}\quad , \quad 
S[\theta]=\int\!\! dt \left[\half\theta^i\id\theta^j\delta_{ij}-V(\theta^i)\right].
\ee

Following the same coarse-graining scheme explained in the previous subsections we modify 
the symplectic structure of the reduced phase space by replacing $\tilde{\sigma}$
with $\tilde{\sigma}_k$ = $(1+r_k) \tilde{\sigma}$.
This is tantamount to add the term 
$\Delta S_k[\theta]=\int\!\! dt \left[\half\theta^i r_k(-\dt^2)\id\theta^j\delta_{ij}\right]$ 
to the bare action.
Correspondingly the functional measure becomes: 
$\mu_k=( {\rm SDet}\, \frac{\tilde{\sigma}_k}{2\pi})^{1/2} = \mu\, 
\left({\rm SDet}(1+r_k)\de\right)^{1/2}$,
where $\de$ stands for a product of Dirac and Kronecker deltas.
Then the modified path integral reads
\be\label{realGrassmannZk}
 Z_k[J_i]=\int [d\theta] \mu_k[\theta] 
 e^{\,i\left\{ S[\theta]+\Delta S_k[\theta]+J_i\cdot\theta^i\right\}}.
\ee
Starting from it, one defines the AEHA
\be
\Delta S_k[\bth]+\Gamma_k [\bth^i]=
\mathop{\min}_{J_i}\left( W_k[J^i] - J_i \cdot \bth^i \right)
\ee
which satisfy the following integro-differential equation
\be
e^{\,i\Gamma_k[\bth^i]}=\int [d\theta] \mu_k[\theta]
 e^{\,i\left\{ S[\theta]+\Delta S_k[\theta-\bth]
-\Gamma_k\frac{\overleftarrow{\de}\phantom{\bth}}{\de\bth^i}\cdot(\theta-\bth)^i
\right\}}
\,.
\ee
and therefore the $k\rightarrow\Lambda$ limit of $\Gamma_k$ is just the bare action. 
The flow equation for $\Gamma_k$ reads
\be
i\dot{\Gamma}_k=-\frac{1}{2}{\rm Tr}\left[\dot{r}_k (1+r_k)^{-1}\de\right]
+\half{\rm Tr}\left[\left(\dot{r}_k\id\de\right)
\left( r_k\id\de+\frac{\overrightarrow{\de}\phantom{\bth}}{\de\bth}\Gamma_k\frac{\overleftarrow{\de}\phantom{\bth}}{\de\bth}
\right)^{-1}
\right]\,.
\ee
where the trace is over $\{i,j\}$ indices as well and, as in the bosonic case,
in the matrix $r_k i\partial\de$ the derivatives act on the first index.

\section{The effective Hamiltonian action in quantum field theory}

There are at least two possible generalizations of the previous formalism 
to quantum field theory (QFT). 

The simplest can be obtained by embracing the traditional Hamiltonian formulation
of field theory, where one associates a canonically conjugate field (momentum)
to the time derivative of each Lagrangian coordinate. This choice leads to a non covariant formulation.
The translation of all previous formulas to this framework can be
obtained by replacing the bare Hamiltonian with the spatial integral of a Hamiltonian density,
and promoting the integrals and functional traces to sums over spatial positions as well
as time instants.
In this way one can obtain a formal definition of the non covariant effective Hamiltonian action and extend
all the previous discussions developed in section $2.1$.

However, in so doing, willing to construct the corresponding coarse-graining procedure for the flow of the
average effective Hamiltonian action, one faces the necessity to regularize the spatial part of these summations,
which are otherwise ill-defined. In other words the regulator matrix $R_k$, appearing
in $\Delta S_k$ and $\mu_k$, must now contain operators depending on spatial derivatives too.
For instance, choosing an off diagonal $R_k$ one could consider
\ba
R_k(x,x')&=&
\begin{pmatrix}
0 & r_k(-\Box)\partial_0 \de(x-x') \\
-r_k(-\Box)\partial_0 \de(x-x') & 0\\
\end{pmatrix}\nonumber\\
\mu_k&=&\left[{\rm Det}\frac{1}{2\pi}
\begin{pmatrix}
0 & \left(1+r_k(-\Box)\right)\partial_0 \de(x-x') \\
-\left(1+r_k(-\Box)\right)\partial_0 \de(x-x') & 0\\
\end{pmatrix}\right]^\half\nonumber
\ea
but this choice would explicitely break Lorentz symmetry.
Instead it would be easy to write more general regulators preserving
such a symmetry, even if in an implicit form.
In both cases one may study the AEHA defined by the integro-differential equation
\be
e^{i \Gamma_k[\bpi,\bphi]}=\int \left[ d\pi d\phi\right]\mu_k[\pi,\phi]  
e^{\, i\left\{S[\pi,\phi]+\Delta S_k[\pi-\bpi,\phi-\bphi]
-(\pi-\bpi)\frac{\de \Gamma_k}{\de \bpi}-(\phi-\bphi)\frac{\de \Gamma_k}{\de \bphi}\right\} }
\nonumber \,.
\ee
This road could be useful if one is interested in non-relativistic field theories,
but for relativistic systems,
since Lorentz invariance is not manifest, in this framework
 it is hard to distinguish truncations for $\Gamma_k$
that are Lorentz symmetric from those that are not
(one would have to deal with Ward-Takahashi-Slavnov-Taylor identities).

Another possibility is to choose a covariant Hamiltonian formalism, in which
one introduces a momentum field for each first order partial derivative of the Lagrangian 
coordinates, thus preserving manifest Lorentz covariance. 
In the following we will give the two simplest examples of how this could work:
spin zero and spin one half field theories.
There are several choices one can do.
In this work we shall attempt to use a {\it reduced} approach, which has the advantage of being the minimal extension,
which on one side preserves the general results in $0\!+\!1$ dimensions
(QM) and on the other side leads to the usual QFT results
 in the case of quadratic bare Hamiltonians.
More general formulations as well as specific applications will be considered elsewhere.

\subsection{Covariant Hamiltonian scalar field theory}
Let us build the covariant Hamiltonian formulation of a classical unconstrained single scalar field in $d$ spacetime dimensions with the standard Lagrangian density
\be
{\cal L}(\phi,\partial_\nu \phi)=-\frac{1}{2}\left(\partial_\nu \phi\right)\left(\partial^\nu \phi\right)-V(\phi)\nonumber
\ee
(in a Minkowski mostly-plus signature).
The covariant Hamiltonian density is defined as the extremum
\be\label{bareH_QFT}
{\cal H}(\pi^\nu,\phi)=\mathop{\rm ext}_{\partial_\nu \phi}\left(-\pi^\nu \partial_\nu \phi- 
{\cal L}(\phi,\partial_\nu \phi) \right)
=-\frac{\pi_\nu \pi^\nu}{2}+V(\phi)
\ee
and by demanding the stationarity of the Hamiltonian action 
\be\label{covariant action}
S=\int\!\!d^dx\, \left[ -\pi^\nu \partial_\nu\phi-{\cal H}\right]
\ee
one finds the De Donder-Weyl equations
\be
\pi^\nu =\partial^\nu\phi \quad , \quad \partial_\nu\pi^\nu =V^\prime(\phi)\nonumber
\ee
i.e. a first order system equivalent to $\Box \phi-V^\prime(\phi)=0\, $.
Here the dynamics of $\pi^\nu$ and $\phi$ seem to be completely coupled, however this is not the case.
In fact the Lorentz vector $\pi^\nu$ can be decomposed into a transverse and a gradient part
$\pi^\nu=\pi^\nu_\perp+\pi^\nu_\parallel$, by means of the standard projectors
$\Pi_\parallel^{\mu\nu}=\partial^\mu\partial^\nu/\Box$
and $\Pi_\perp^{\mu\nu}=\eta^{\mu\nu}-\Pi_\parallel^{\mu\nu}$.
Rewriting the Hamiltonian action density in terms of these
reduced degrees of freedom (and assuming that the boundary terms coming
from integration by parts do not contribute)
one finds $-\pi_\parallel^\nu\partial_\nu\phi-{\cal H}$ 
with
\be
{\cal H}(\pi_\perp^\nu,\pi_\parallel^\nu,\phi)=-\frac{\pi_{\perp\nu} \pi_\perp^\nu}{2}-
\frac{\pi_\parallel^\nu \pi_{\parallel\nu}}{2}+V(\phi)\nonumber
\ee
and the corresponding Hamiltonian equations
\be
\partial_\nu\pi_\parallel^\nu =\Box\phi \quad , \quad 
\partial_\nu\pi_\parallel^\nu=V^\prime(\phi) \quad , \quad \pi_\perp^\nu=0\nonumber\, .
\ee
Hence the transverse momenta are classically irrelevant if the Hamiltonian is quadratically depending on them. 
This translates into the following quantum property: if the bare Hamiltonian is separable in $\pi$ and $\phi$
and quadratically depending on $\pi$, the functional integration over transverse momenta factorizes from those on the other two fields.

Now let us address the possibility to extend this formalism to covariant
 Hamiltonian densities that are more than quadratic in the momenta.
The classical decoupling of the transverse momenta, i.e. their factorization in the functional integral, can happen also for
non-quadratic Hamiltonians, such as for instance
${\cal H}=T(\pi^\mu \pi_\mu)+V(\phi)$.
Insisting in the validity of the classical variational principle for the action~\eqref{covariant action}
the classical equations read
\be
\partial_\nu\pi_\parallel^\nu ={\de\;\over\de\phi}\int\!\! d^dx{\cal H} 
\quad , \quad 
-\partial_\nu\phi={\de\phantom{\pi}\over\de\pi_\parallel^\nu}\int\!\! d^dx{\cal H} 
\quad , \quad 
{\delta\phantom{\pi}\over\delta \pi_\perp^\nu}\int\!\! d^dx{\cal H}=0\nonumber\, .
\ee
The interesting question now is whether the third equation is a constraint or it gives a dynamics to the transverse momenta.
If ${\cal H}$ does not contain derivatives of $\pi_\perp^\nu$, and if one can perform some sort of Fourier transform
such that $\pi_\perp^\nu$ can be considered orthogonal to $\partial^\nu$ with respect to the metric in Minkowski
spacetime, then the third equation cannot contain derivatives of $\pi_\perp^\nu$.
Therefore, under these assumptions, one can always solve the third equation by writing
$\pi_\perp^\nu$ as a local (if ${\cal H}$  is local) function of $\pi_\parallel^\nu$,
 $\phi$ and their derivatives.
By substituting this solution in the first two equations one gets a coupled dynamics for 
the unconstrained variables $\pi_\parallel^\nu$ and $\phi$ only.
That is, under these assumptions the transverse momenta do not have their own independent dynamics
and behave only as redundant variables which can be eliminated without
 loosing the locality of the action.
However even in this case the quantization of the theory containing the $\pi_\perp^\nu$ fields
is not equivalent to the quantization of the theory in which one got rid of them by means
of the classical equations, since in the first case one has a full functional integral
over $\pi_\perp^\nu$, whose stationary phase approximation gives the second quantum theory.
Nevertheless, considering an Hamiltonian action depending on parallel momenta
only, although it is not the most general case, is already a consistent 
and covariant generalization of the standard non-covariant Hamiltonian
 approach, reproducing the known results for quadratic Hamiltonians.
Therefore in this paper we will restrict ourselves to such a case.

The aim of the rest of this section is to give meaning to the quantization
of the classical theory with the bare action~\eqref{covariant action} under the
assumption that ${\cal H}$ depends on $\pi_\parallel^\nu$ only.
Since in this case the bare action $S$ does not depend on $\pi_\perp^\nu$,
we are in presence of a gauge symmetry: by introducing projectors where
needed, $S$ can be rewritten in a form which is manifestly invariant under the infinitesimal
transformation: $\de\pi^\nu(x)=\Pi_\perp^{\nu\rho}\epsilon_\rho(x)$, for any infinitesimal
vector field $\epsilon$. In this paper we will discuss
the functional integral quantization of the theory by means of the
introduction of the constraint $\Pi_\perp^{\nu\rho}\pi_\rho=0$ in the functional measure
(something like a sharp gauge fixing\footnote{Dirac's classification of constraints
and the consequent quantization schemes for gauge theories are based on the
non-covariant Hamiltonian formalism and therefore are not straightforwardly
applicable to the present case. However classical 
constrained dynamics has been extensively discussed in the literature
 about the covariant Hamiltonian formalism(s)~\cite{GMS}
and some proposals have been provided about the corresponding path
integral quantization of gauge theories~\cite{BS}.}).
Thus, the generating functional of the theory will be
\be\label{ZQFT}
Z[I_\nu,J]=e^{i W[I_\nu,J]}=\int \left[d\pi^\nu d\phi\right]\de\!\left[\Pi_\perp^{\nu\rho}\pi_\rho\right] 
\mu\, e^{\,i\left\{S[\pi^\nu,\phi]+I_\nu\cdot\pi^\nu+J\cdot\phi\right\} }\,.
\ee
Notice that, depending on which regularization and precise definition of the
functional integral is chosen, the functional integration
over $[d\pi_\parallel^\nu]$ and the constrained integration 
$[d\pi^\nu]\de\!\left[\Pi_\perp^{\nu\rho}\pi_\rho\right]$
could differ by a field-independent Jacobian determinant.
A skeletonized definition in Fourier space, i.e. the use of a discretization 
of Fourier space, would make this Jacobian to be equal to one.
Whenever such a Jacobian is unity, 
since the constraint kills all but one of the integrals over the $\pi$'s,
the usual functional measure $\mu={\rm Det}\frac{1}{2\pi}$ 
provides the normalization needed
in order to reproduce the known results for bare Hamiltonian actions
quadratic in the momenta. Otherwise $\mu$ needs to be different (but still field-independent)
in order to balance the Jacobian determinant.
Starting from eq.~\eqref{ZQFT} the definition of the effective Hamiltonian action is again 
\be\label{defGamma1QFT}
\Gamma\left[\bpi^\nu,\bphi\right] =
\mathop{\rm ext}_{I_\nu,J}\left( W[I_\nu,J] -I_\nu \cdot\bpi^\nu -J\cdot \bphi\right)
\ee
which is equivalent to state that $\Gamma$ is the solution of the following integro-differential equation with suitable boundary conditions
\be\label{intdiffeqQFT}
e^{i \Gamma[\bpi^\nu,\bphi]}=\int \left[ d\pi^\nu d\phi\right]
\de\!\left[\Pi_\perp^{\nu\rho}\pi_\rho\right]\mu \, 
e^{\, i\left\{S[\pi^\nu,\phi]
-(\pi-\bpi)^\nu\frac{\de \Gamma}{\de \bpi^\nu}-(\phi-\bphi)\frac{\de \Gamma}{\de \bphi}\right\} }\,.\\
\ee

In the following we shall try to give a definition of the
integrals~\eqref{ZQFT} and~\eqref{intdiffeqQFT} based on an RG flow equation for the average version of the effective action.
First of all, one has to introduce $k$-dependent operators that disappear in the
$k\rightarrow0$ limit and that provide a rising delta functional 
in the $k\rightarrow \Lambda$ limit.
As before let us denote this regularization as follows
\be
Z_k[I_\nu,J]=\int \left[d\pi^\nu d\phi\right]\de\!\left[\Pi_\perp^{\nu\rho}\pi_\rho\right]
\mu_k\,
e^{\,i\left\{ S[\pi^\nu,\phi]+\Delta S_k[\pi^\nu,\phi]+I_\nu\cdot\pi^\nu+J\cdot\phi\right\} } \, .\nonumber
\ee
We will choose a regularization corresponding to a $k$-dependent deformation
of the term whose one-dimensional version is the Legendre transform term, i.e.
$-\pi^\mu \partial_\mu \phi$. In other words, we will restrict to an off-diagonal $R_k$, or
more explicitly
\ba\label{deltaSkQFT}
\Delta S_k[\pi^\nu,\phi]&=&\int\!\!d^dx\left[-\pi^\nu r_k(-\Box)\partial_\nu\phi\right]\\
\label{mukQFT}
\mu_k&=&\mu\left[{\rm Det}
\begin{pmatrix}
0 & -\left(1+r_k(-\Box)\right)\partial_\nu \de(x-x') \\
\left(1+r_k(-\Box)\right)\partial_\nu \de(x-x') & 0\\
\end{pmatrix}\right]^\half\nonumber\, .
\ea
The definition of the AEHA is the same as in quantum mechanics
\be\label{defGammasQFT}
\Gamma_k\left[\bpi^\nu,\bphi\right] +\Delta S_k\left[\bpi^\nu,\bphi\right]=
\mathop{\rm ext}_{I_\nu,J}\left( W_k[I_\nu,J] -I_\nu \cdot\bpi^\nu -J\cdot \bphi\right)
\ee
wherefrom the usual integro-differential equation
\be\label{intdiffeqkQFT}
e^{i \Gamma_k[\bpi^\nu,\bphi]}=\int \left[ d\pi^\nu d\phi\right]
\de\!\left[\Pi_\perp^{\nu\rho}\pi_\rho\right]\mu_k \, 
e^{\, i\left\{S[\pi^\nu,\phi]+\Delta S_k[(\pi-\bpi)^\nu,\phi-\bphi]
-(\pi-\bpi)^\nu\frac{\de \Gamma_k}{\de \bpi^\nu}-(\phi-\bphi)\frac{\de \Gamma_k}{\de \bphi}\right\} }\,.\\
\ee
By taking the $k\partial_k$ derivative of eq.~\eqref{intdiffeqkQFT} one finds
\be\label{pre_flow_eq_manypi}
i\dot{\Gamma}_k=\frac{\dot{\mu}_k}{\mu_k}-i\int\!\!d^dx
\langle(\pi-\bpi)^\nu \dot{r}_k\partial_\nu(\phi-\bphi)\rangle\,.
\ee
For the second term, we need to write the two point function in terms of derivatives
of $\Gamma_k$.
Since this theory contains one Lagrangian coordinate and one momentum,
$\Gamma^{(2)}$ is a two-dimensional square matrix, as in quantum mechanics.
However, our momentum is a vector field bringing a Lorentz index, and even if it lies
in a one-dimensional subspace, such a subspace varies from point to point in spacetime.
Thus, unless we want to choose a frame in the tangent bundle such that at every spacetime point
$x$ the vector $\pi^\nu(x)$ has only one and the same non-vanishing component, we are forced to
deal with it as a generic Lorentz vector. Since we prefer to write formulas in a generic frame,
we will treat $\Gamma^{(2)}$ as a generic $(d+1)$-dimensional square matrix,
whose entries can be written as four blocks: a $(1,1)$ tensor ($d$-by-$d$ square matrix),
one contravariant (column) vector, one covariant (row) vector, and one Lorentz scalar.
Because the momenta enter the theory naturally with high indices (to be contracted with derivatives), we will treat them as column vectors.
Therefore the source $I$ will become a row vector. 
We will denote by $()^t$ the transposition of these objects, that is the canonical isomorphism defined by the spacetime metric.
Thus $\pi^t$ and $I^t$ will denote row and column vectors respectively.
Of course derivatives with respect to contravariant (covariant)
vectors will be considered  covariant (contravariant). 
Going back to the task of computing the two point functions, since
\be
i\langle {\cal T}\!
\begin{pmatrix}
(\pi\!-\!\bpi)_x\otimes(\pi\!-\!\bpi)^t_{x'}&(\pi\!-\!\bpi)_x(\phi\!-\!\bphi)_{x'}\\
(\phi\!-\!\bphi)_x(\pi\!-\!\bpi)_{x'}^t&(\phi\!-\!\bphi)_x(\phi\!-\!\bphi)_{x'}\\
\end{pmatrix}
\rangle_k 
= W^{(2)}_{\!k\phantom{(2}x x'}[I,J]=\!
\begin{pmatrix}
\frac{\delta W_k}{\delta I_x}\otimes\big(\frac{\overleftarrow{\de}\;\;}{\delta I_{x'}}\big)^t&
\frac{\delta^2W_k}{\delta J_{x'}\delta I_x}\\
\big(\frac{\delta^2W_k}{\delta I_{x'}\delta J_x}\big)^t&
\frac{\delta^2W_k}{\delta J_{x'}\delta J_x}\\
\end{pmatrix}\nonumber
\ee
one needs an explicit expression for the vector
$\frac{\delta^2W_k}{\delta J \delta I}$
in terms of $\Gamma_k$.
This can be found by using
\be
I_\mu= r_k\partial_\mu\bphi-\frac{\de \Gamma_k}{\de \bpi^\mu} \quad , \quad
J=- r_k\partial_\nu\bpi^\nu-\frac{\de \Gamma_k}{\de \bphi}\nonumber
\ee
thus getting
\ba
 &&W^{(2)}_{\!k\phantom{(2}x x'}[I,J]=
\begin{pmatrix}
\bpi\otimes\big(\frac{\overleftarrow{\de}\;}{\de I}\big)^t&\frac{\delta\bpi}{\delta J}\\
\big(\frac{\delta \bphi}{\delta I}\big)^t&\frac{\delta \bphi}{\delta J}\\
\end{pmatrix}_{x x'}=
\begin{pmatrix}
I^t\otimes\frac{\overleftarrow{\de}\;}{\delta \bpi}&
\big(\frac{\delta I}{\delta \bphi}\big)^t\\
\frac{\delta J}{\delta \bpi}&\frac{\delta J}{\delta \bphi}\\
\end{pmatrix}^{-1}_{x x'}\nonumber\\
&=&-\begin{pmatrix}
\big(\frac{\de \Gamma_k}{\de\bpi}\big)^t\otimes\frac{\overleftarrow{\de}\;}{\de\bpi}&
\big(\!-\!r_k\partial\de+\frac{\de^2 \Gamma_k}{\de \bphi\de\bpi}\big)^t\\
r_k\partial\de+\frac{\de^2 \Gamma_k}{\de \bpi\de\bphi}&
\frac{\de^2 \Gamma_k}{\de \bphi\de\bphi}\\
\end{pmatrix}^{-1}_{x x'}
\equiv-\begin{pmatrix}
A&B\\
B^T&D\\
\end{pmatrix}^{-1}_{x x'} \nonumber\\
\nonumber
\ea
where $(r_k\partial\de)_{x_1 x_2}=r_k(-\partial_{x_1}^2)\partial_{x_1}\de(x_1-x_2)$ is a Lorentz covariant (row) vector.
This matrix is manifestly symmetric with respect to full transposition $T$ of both Lorentz and spacetime-position indeces.
Since the building blocks $B$ and $B^T$ are not square matrices, we cannot use formula~\eqref{blockinversion}.
Anyway, if $A$ and $(D-B^TA^{-1}B)$ are non singular this becomes
\be\label{nonsquare_inversion1}
 W^{(2)}_{\!k}[I,J]=-
\begin{pmatrix}
A^{-1}+A^{-1}B(D-B^TA^{-1}B)^{-1}B^TA^{-1} & -A^{-1}B(D-B^TA^{-1}B)^{-1}\\
-(D-B^TA^{-1}B)^{-1}B^TA^{-1} & (D-B^TA^{-1}B)^{-1}\\
\end{pmatrix}
\ee
if instead $D$ and $(A-BD^{-1}B^T)$ are non singular, then we can write
\be\label{nonsquare_inversion2}
 W^{(2)}_{\!k}[I,J]=-
\begin{pmatrix}
(A-BD^{-1}B^T)^{-1}&-(A-BD^{-1}B^T)^{-1}BD^{-1}\\
-D^{-1}B^T(A-BD^{-1}B^T)^{-1}&D^{-1}+D^{-1}B^T(A-BD^{-1}B^T)^{-1}BD^{-1}\\
\end{pmatrix}\,.
\ee
The off-diagonal entries of these matrices can be finally plugged into 
eq.~\eqref{pre_flow_eq_manypi}.
Thus, if for instance $A$ and $(D-B^TA^{-1}B)$ are non singular the final flow equation is
\ba
i\dot{\Gamma}_k&=&{\rm Tr}\left[\dot{r}_k\left(1+r_k\right)^{-1}\de\right]
-{\rm Tr}\Bigg[
\left(r_k\partial\de+\frac{\de^2 \Gamma_k}{\de \bpi\de\bphi}\right)
\left(\frac{\de^2\Gamma_k}{\de\bpi\de\bpi}\right)^{-1}
\left(\dot{r}_k\partial\de\right) \nonumber\\
& &\left[\frac{\de^2 \Gamma_k}{\de \bphi\de\bphi}-
\left(r_k\partial\de+\frac{\de^2 \Gamma_k}{\de \bpi\de\bphi}\right)
\left(\frac{\de^2 \Gamma_k}{\de\bpi\de\bpi}\right)^{-1}
\left(r_k\partial\de+\frac{\de^2 \Gamma_k}{\de \bpi\de\bphi}\right)^T\right]^{-1}
\Bigg]\,.
\ea
Here for sake of notational simplicity we dropped the symbols for tensor products and Lorentz
transpositions.
By means of eq.~\eqref{nonsquare_inversion2} the reader can write down a similar flow equation for the case in which $D$ and $(A-BD^{-1}B^T)$ are non singular.

As an example let's discuss the LHA for a scalar theory enjoying
$Z_2$-symmetry under simultaneous reflections: $\pi^\nu\to-\pi^\nu$, $\phi\to-\phi$. 
In other words, we are going to insert the approximation
$\Gamma_k=\int d^dx\big(-\bpi^\nu\partial_\nu\bphi-{\cal H}_k(\frac{\bpi^2}{2},\frac{\bphi^2}{2})\big)$, where $\bpi^2\equiv\bpi^\nu\bpi_\nu$, in the previous flow equation. 
In order to project the r.h.s. of the flow equation inside such an ansatz
for $\Gamma_k$, one usually evaluates it on constant field configurations. This can be done also in the present case, without
contradicting the assumption that the momenta $\bpi^\nu$ be longitudinal, by choosing the Fourier transform of $\bpi^\nu$ pointing
in the same direction of the Fourier variable and being
proportional to a delta function.
We will denote by $\cH_k^{(i,j)}$ the
result of differentiating $\cH_k$ $i$-times w.r.t. $\frac{\bpi^2}{2}$
and $j$-times w.r.t. $\frac{\phi^2}{2}$.
Let us recall the notation already used in quantum mechanics 
(see eq.~\eqref{lha_flow_P}) for the regulator in the LHA, i.e.
$P_k(-\Box)=(1+r_k(-\Box))^2(-\Box)$. Let us also introduce for convenience
the function
\be\label{sigma_d}
\sigma_d(\alpha)=\!\!\!\!\phantom{F}_2F_1\left(\half,1;\frac{d}{2};\alpha\right)
\ee
and the following threshold functional
\be
l^d_0[\alpha,\beta]=\frac{1}{4}v_d^{-1}k^{-d}\int\frac{d^dp}{(2\pi)^d}
\frac{\dot{P}_k(p^2)}{P_k(p^2)+k^2\beta(p^2)}
\sigma_d(\alpha(p^2))
\ee
where $v_d^{-1}=2^{d+1}\pi^{d/2}\Gamma(\frac{d}{2})$.
Then the flow equation for the dimensionful average effective Hamiltonian
density can be written
\be\label{QFTLHA}
i\dot{\cH}_k=2v_d k^d\left(
l^d_0[\alpha_\cH,\beta_\cH]-l^d_0[\alpha_\cH,0]\right)
\ee
where we further defined the dimensionless quantities
\ba\label{general_alpha}
\alpha_\cH(p^2)&=&\frac{P_k(p^2)}{P_k(p^2)+k^2\beta_\cH}
\frac{\bpi^2\cH_k^{(2,0)}}{\cH_k^{(1,0)}+\bpi^2\cH_k^{(2,0)}}\\
\label{beta}
\beta_\cH&=&\frac{1}{k^2}\left[\bpi^2\bphi^2\left(\cH_k^{(1,1)}\right)^2
\frac{\cH_k^{(1,0)}}{\cH_k^{(1,0)}+\bpi^2\cH_k^{(2,0)}}
-\cH_k^{(1,0)}
\left(\cH_k^{(0,1)}+\bphi^2\cH_k^{(0,2)}\right)\right]
\ea
the second of which is not a function of $p^2$.
First of all let us notice that if we make the ansatz that 
the theory be quadratic in the momenta at every scale, then the
vanishing of $\cH_k^{(2,0)}$ entails the vanishing of $\alpha_\cH$
and we recover the Lagrangian flow in the LPA.
If instead $\alpha_\cH$ is non-vanishing, the presence of a 
$p$-dependent denominator in the argument of the function $\sigma_d$
in general makes the analytic computation of $l^d_0$ quite hard.
For this reason it is wise to choose the regulator in such a way to kill
the $p$-dependence of all the denominators. In the LHA this can be
accomplished by means of the optimized regulator 
$r_k(p^2)=\big(k/\sqrt{p^2}-1\big)\theta(k^2-p^2)$, 
i.e. $P_k(p^2)=(k^2-p^2)\theta(k^2-p^2)$.
For such a choice
\be\label{Litim_alpha}
\alpha_\cH(p^2)=\frac{1}{1+\beta_\cH}
\frac{\bpi^2\cH_k^{(2,0)}}{\cH_k^{(1,0)}+\bpi^2\cH_k^{(2,0)}}
\ee
is $p$-independent and the threshold function for constant
argument becomes
\be
l^d_0[\alpha,\beta]=\frac{2}{d}\frac{1}{1+\beta}
\sigma_d(\alpha)\nonumber\,.
\ee
To sum up, for the optimized regulator the flow equation of the LHA
reads (after Wick rotation)
\be\label{QFTLHA_Litim}
\dot{\cH}=-\frac{4}{d}v_d k^d \frac{\beta_\cH}{1+\beta_\cH}
\sigma_d(\alpha_\cH)
\ee
with $\beta_\cH$ and $\alpha_\cH$ given by~\eqref{beta} 
and~\eqref{Litim_alpha}.
The function $\sigma_d$ takes simpler forms for integer $d$.
For instance, in $d=2$, $d=3$ and $d=4$ it respectively reads
\be
\sigma_2(\alpha)=(1-\alpha)^{-\half}\quad , \quad 
\sigma_3(\alpha)=\frac{{\rm arctanh}(\sqrt{\alpha})}{\sqrt{\alpha}}
\quad , \quad 
\sigma_4(\alpha)=\frac{2}{\alpha}\left[1-(1-\alpha)^{-\half}\right]\,.
\ee
Equation~\eqref{QFTLHA} can be taken as a first step towards the
nonperturbative study of scalar QFT in the covariant
Hamiltonian formalism. In particular, one of the first questions to be
addressed is whether such an equation admits non-Gaussian fixed points.
In case a positive answer exists, these could provide a possible
 solution to the triviality problem of scalar QFT in four dimensions.
In fact, choosing the engineering dimensions of the
fields in such a way that the coefficients of the $\bpi^2$ and 
Legendre terms be dimensionless, dimensional analysis tells us that the
coupling multiplying the operator $(\bpi^2)^i(\bphi^2)^j$ has dimensionality
$d_{ij}=(1-i-j)d+2j$. Therefore in $d=4$ the only momentum dependent non IR-irrelevant
term is $\bpi^2$, all other terms with positive integers $(i,j)$ being IR-irrelevant.
In other words, scalar theories more than quadratic in the momenta are expected to be
highly favored in the UV and to be well approximated by quadratic
theories in the IR. From this point of view it seems reasonable to
look for the UV completion of four dimensional scalar QFT in a general
Hamiltonian framework. For instance this could be done according
to the paradigm of asymptotic safety~\cite{WeinbergAS}, i.e.
by looking for nontrivial fixed points of the RG flow having
a finite dimensional UV critical surface (a finite number of
UV attractive directions in theory space).
On the other hand this very same argument in the case of a simpler scalar QFT
in configuration space is often used for a qualitative understanding
of the absence of $Z_2$-symmetric non-Gaussian fixed points in $d=4$: 
in this case the only IR-relevant monomial-like operator is the mass
term, all other monomials being either marginal or IR-irrelevant.
 Anyways in the present formulation the theory contains
not only a scalar field but also a longitudinal vector field, therefore
we believe that the understanding of this issue requires explicit computations in order to reveal the details of the underlying dynamics.

Another interesting question regarding eq.~\eqref{QFTLHA} is
whether it can teach us to what extent the covariant Hamiltonian framework adopted in this paper is sound and useful. In particular, it would be interesting to compare, within a fixed approximation
such as the LHA, the RG flow of the traditional
non-covariant Hamiltonian formulation with that of the covariant one allowing for longitudinal momenta only (the present case) 
and with the one allowing also for transverse momenta.
These and other questions will be left open by the present work.

\subsection{Spinor field theory}
Let us build the covariant Hamiltonian formulation of a classical Lagrangian field theory 
for a single Dirac field in a number $d$ (allowing Dirac spinors) of spacetime dimensions with the standard Lagrangian density
\be
{\cal L}(\psi,\partial_\nu \psi)=-\bpsi\cin\psi-V(\bpsi,\psi)\nonumber
\ee
(in a Minkowski mostly-plus signature) where $\bpsi=i\psid\gamma^0$ .
Defining the momenta $\pi^\nu$ as the right partial derivatives of 
$-{\cal L}$ with respect to $\partial_\nu\psi$
we find $d$ second class primary constraints:
\be\label{QFTconstraints}
\chi^\nu(x) = \pi^\nu(x)-\bpsi\gamma^\nu(x)=0 
\ee
whose solution is $\bpsi=\frac{1}{d}\pi^\nu\gamma_\nu$.
These constraints boil down the momenta to functions of just one field,
hence there is no room here for the other $d-1$ conjugate fields that in the bosonic case
could be identified with the transverse momenta.
The relevant phase space is the surface $\cal{S}$ defined by ($\ref{QFTconstraints}$), 
the only independent coordinate on it is $\psi$ and the functional integral is to be taken over 
all histories $\psi(x)$.
The covariant Hamiltonian density is defined as
\be
{\cal H}(\pi^\nu,\psi)=\mathop{\rm ext}_{\partial_\nu \psi}
\left(-\pi^\nu \partial_\nu \psi- {\cal L}(\psi,\partial_\nu \psi) \right)
=V(-\frac{1}{d}\pi^\nu\gamma_\nu,\psi)\nonumber
\ee
and on $\cal{S}$ it is just $V(\bpsi,\psi)$.
Thus the covariant AEHA formalism in this case is equivalent to the usual Lagrangian approach, 
exactly as was previously described for fermionic QM, one has just to replace time derivatives with $\cin$ operators.

\section{Conclusions}
In this work we have focused on the description of quantum dynamics by means of the quantum effective Hamiltonian action (EHA).
We have first reviewed its properties by a discussion in quantum mechanics,
taking advantage of the fact that QM and non covariant QFT's are very similar in this respect.
We have then discussed how to compute the effective action.
For instance we have derived a general one loop formula, which can be useful to compare the results obtained by other approaches, and we have generalized
the variational definition provided a long time ago by 
Jackiw and Kerman~\cite{JK} for its Lagrangian counterpart. 
But the main goal of this work is to provide an alternative 
non-perturbative tool to compute the EHA. 
This is an Hamiltonian generalization of the so-called functional renormalization group, in particular of the formulation
due to Wetterich based on the average effective (Lagrangian) 
action~\cite{wetterich}.

Such a generalization, which is one of the main results of our work, is straightforward in QM, even if the 
one-parameter-dependent family of cutoff operators is wider and in general the formulae are more cumbersome.
Starting from the most general flow equation we have derived simpler equations like the one associated to the so called
local Hamiltonian approximation (LHA), i.e. the leading order in the derivative expansion.
In order to show that the approach is trustworthy, we have studied, as an example, a family of quantum mechanical
systems with bare Hamiltonians non quadratic in the momenta, we have computed for two cases the ground state
energy and the first energy gap, and we have successfully compared them to the exact results,
employing different kind of schemes and approximations.
We stress that for the models under consideration we needed to take into account, as expected,
the issue of Weyl ordering, which turns out to be at the base of the present flow equation quantization
as it is well known to be for the functional integral quantization. This fact calls for some care in defining the concept
of a bare non separable in phase space Hamiltonian action.

The application of the formalism developed for QM to the QFT case is
straightforward and quickly discussed but, as in all Hamiltonian
approaches to QFT, one must pay full generality and manifest unitarity
with non-manifest Lorentz covariance. This is unpleasant and complicates
the job of performing approximations without breaking such a symmetry.
For this reason, in the second part of the paper, we have discussed the
possibility to generalize the EHA formalism to include also
covariant Hamiltonian QFT.
Functional integral quantizations of such theories has already been
addressed in the literature, especially for gauge theories.
In the present work we have addressed the simplest cases of scalar
and spinor degrees of freedom. Actually, for scalar QFT we further
restricted our work to the presence of one conjugate momentum only,
namely a longitudinal vector field. In this specific case we have
provided an RG flow equation representation of the corresponding
QFT, and we have worked out its explicit form in the LHA.

Let us close this work addressing the issue of the
physical motivations for it and of its usefulness.
Clearly, the use of this framework is related to Hamiltonian systems
non quadratic in momenta, therefore we should comment on the question:
where are them or why should we look for them?

Quantum mechanical systems more than quadratic in the
momenta may be interesting on the base of first principles
(think about the action of the free relativistic particle)
or arise as effective descriptions of physical systems.
Also, they could appear as intermediate technical tools for the
description of more complicated systems. For instance,
within the worldline formalism, one-loop computations 
are reduced to quantum mechanical path integrals
with Hamiltonians which sometimes are
 non-quadratic in the momenta~\cite{worldline}. 
In these cases one can hope to use this approach as an alternative or a complementary tool to perturbation theory.

Theories more than quadratic in the momenta, when reduced to the Lagrangian formulation, show a nonlinear dependence on the derivatives of the fields. This dependence, if expanded in powers and truncated, typically generates violations of unitarity. 
Nevertheless before truncation nothing prevents such theories from
being unitary. That is, there might be some interesting non trivial extensions of quantum models which are non-quadratic in the momenta and 
that make perfectly sense from a quantum mechanical point of view.

Why should we look for them? As already commented at the end of the
section on scalar QFT, the study of the RG flow on the Hamiltonian theory space
might show new possibilities for the UV or IR behavior of systems
that at some intermediate scale are well approximated by simple
Lagrangian theories. Stated in different words, keeping
both phase space variables could make easier the task of parameterizing
the quantum dynamics far from that intermediate simple Lagrangian scale.
One reason for such an expectation is the following: we know that the
effective actions are in general non-local, and that integrating out
non-Gaussian degrees of freedom is responsible for such non-localities, therefore avoiding to integrate out the momenta should be of help in the
 hard task of reducing as far as possible the importance of non-local
interactions. Restated one more time: even by studying the running of approximate local actions on the Hamiltonian theory space one can, just by putting the
momenta on-shell, have access to at least part of the running of non-local actions in the Lagrangian theory space.
For these reasons also the study of theories whose
bare actions are quadratic but that flow to AEHA's more than quadratic
in the momenta could benefit from this first order formulation.
Examples are the covariant Hamiltonian formulation of
Yang-Mills theory and generic nonlinear sigma models, which 
in our opinion deserve future investigations within the present framework.

The analysis of Hamiltonian flows might open the intriguing 
possibility of finding systems belonging to new universality classes, by looking
for fixed points of the flow in the Hamiltonian formulation.
We have started to consider this challenging problem within the
``reduced"  covariant formulation of scalar QFT presented in this paper,
and we hope to report on this soon.
The results of all these studies will in general depend on the kind 
of Hamiltonian formulation we choose, a fact that enables one to
quantitatively compare different quantization prescriptions
as well as to look for physical systems described by each of them.
Thus, in our opinion, a vast playground lies open, waiting for future investigations.

\vskip 0.2cm
\noindent {\bf Acknowledgments}

We thank Raphael Flore for discussions.

\section{Appendix A: The effective Hamiltonian action as the generating functional of 1PI vertex functions.}
In this appendix we are going to prove that the effective Hamiltonian action is 
the generating functional of the one particle irreducible (1PI) proper vertices,
in the sense that the tree level amplitudes
computed with vertex functions and propagators extracted from it are equal to 
the full perturbative series generated by the bare Hamiltonian action. 
For the ease of the explanation we limit this discussion to the QM case, choosing $\hbar=1$ as a unit of action.
The proof works just as for the usual Lagrangian effective action~\cite{wc}.

1. Write down a path integral based on a Hamiltonian bare action which is $(1/g)$-times the Hamiltonian effective action, 
with $g$ an external parameter. 
This rescaling of the action entails a corresponding rescaling of the 
Liouville form $\lambda_g\equiv\frac{1}{g}\lambda =\frac{1}{g} \bp d\bq $.
Thus, in order to define the new path integral we must adopt a functional measure 
$\mu_g=\sqrt{{\rm Det} \sigma_g}$ corresponding to the symplectic structure $\sigma_g=d\lambda_g$:
\be
e^{iW_g[I,J]}=\int \left[ dp dq\right]\mu_g[p,q] 
e^{\frac{i}{g}\left\{\Gamma[\bp,\bq]+I\cdot\bp+J\cdot\bq\right\}}\ .
\ee

2. Recognize that the parameter $g$ allows one to distinguish different loop
orders in the perturbative evaluation of this path integral.
In fact eqs.(\ref{vertex},\ref{propagatorGamma}) show that in the perturbation theory
generated by $\Gamma_g\equiv\frac{1}{g}\Gamma$ the vertex functions are proportional to $1/g$
while propagators are proportional to $g$. 
Thus any graph with $i$ internal lines and $v$
vertices gives a contribution proportional to $g^{i-v}$.
Since the number of loops is $l=i-v+1$, any loop expansion is an expansion in powers of $g$
of the kind
\be\label{proof3}
W_g[I,J]=\sum_{l=0}^\infty g^{l-1} W_{g,l}[I,J]\,.
\ee

3. Evaluate the same path integral by a stationary phase method, an approximation
	that can be made arbitrarily good by tuning $g$ arbitrarily close to zero.
Since by definition the exponent at the stationarity point gives the $W[I,J]$ of eqn.~\eqref{legendre},
one gets
\be\label{proof4}
e^{iW_g[I,J]}\mathop{\sim}_{g\to0} \ \mu_g[p,q] \left({\rm Det} 
\frac{1}{2\pi g}\Gamma\left[\bp,\bq \right]^{(2)}\right)^{-\half}
e^{\frac{i}{g}W[I,J]}\,.
\ee

4. Expand the logarithm of the last result in powers of $g$. Because
\ba
\log\mu_g[p,q]&=&- {\rm Tr}\log g + \log\mu[p,q] \nonumber \\
\log\left({\rm Det} \frac{1}{2\pi g}\Gamma\left[\bp,\bq \right]^{(2)}\right)^{-\half}\!\!&=&\!\!
 {\rm Tr}\log g + \log\left({\rm Det}\,\frac{1}{2\pi}\Gamma\left[\bp,\bq \right]^{(2)}\right)^{-\half}\nonumber
\ea
the combination of eqs.~\eqref{proof3} and \eqref{proof4} gives
\be
\sum_{l=0}^\infty g^{l-1} W_{g,l}[I,J]\mathop{\sim}_{g\to0} \ 
\frac{1}{g}W[I,J]-i\log\left\{
 \mu[p,q]\left({\rm Det}\,\frac{1}{2\pi}\Gamma\left[\bp,\bq \right]^{(2)}\right)^{-\half}
\right\}
\nonumber
\ee
that is: $W_{g,0}[I,J]=W[I,J]$.

\section{Appendix B: The effective Hamiltonian action from a variational formula on the Hilbert space.}
This appendix is to prove the proposition of section 2 about the possibility to define the
 effective Hamiltonian action in the operator representation by means of a variational principle.
The following arguments are not original, but just the obvious extension of those presented in~\cite{JK}.
In order to compute the extremum~\eqref{variational def} with the constraints~\eqref{constraints psi}
one introduces three Lagrange multipliers $w(t), I(t), J(t)$ and looks for the extremum of
$\langle \psi_- ,t| i\dt -\hat H+J(t)\hat q+I(t)\hat p -w(t)|\psi_+ ,t\rangle$
with respect to the two states $|\psi_\pm,t\rangle$. Setting the two functional derivatives to zero gives
\ba
\left(i\dt -\hat H+J(t)\hat q+I(t)\hat p\right)|\psi_+ ,t\rangle&=&w(t)|\psi_+,t\rangle \label{multipliers}\\
\left(i\dt -\hat H+J(t)\hat q+I(t)\hat p\right)|\psi_- ,t\rangle&=&w^*(t)|\psi_-,t\rangle \,.
\ea
It is possible to define the states
\be
 |+,t\rangle=\exp\left\{i\int_{-\infty}^t\!\!dt'\, w(t')\right\}|\psi_+ ,t\rangle \quad , \quad
 |-,t\rangle=\exp\left\{-i\int_t^{+\infty}\!\!dt'\, w^*(t')\right\}|\psi_- ,t\rangle
\ee
which solve the following Schr\"odinger equation
\ba
\left(i\dt -\hat H+J(t)\hat q+I(t)\hat p\right)|\pm ,t\rangle=0
\ea
and satisfy the boundary conditions: $\mathop{\lim}_{t\rightarrow\mp\infty}| \pm ,t\rangle=|0\rangle\, .$
In other words, 
$|+,t\rangle=\hat U_{I,J}(t,-\infty)|0\rangle$ and $\langle-,t|=\langle0|\hat U_{I,J}(+\infty,t)\,$,
such that 
\be
e^{i W[I,J]}=\langle0|\hat U_{I,J}(+\infty,-\infty)|0\rangle=\langle-,t|+,t\rangle=e^{\,i \int_{-\infty}^{+\infty}\!\!dt'\, w(t')}\, ,
\ee
that is: $W[I,J]=\int_{-\infty}^{+\infty}\!\!dt'\, w(t')\,$.
On the other hand, by contracting eq.~\eqref{multipliers} with $\langle\psi_-,t|$ and using the previous 
equation, along with the constraints~\eqref{constraints psi}, one finds that for the stationarity states
the following relation holds
\be
\int_{-\infty}^{+\infty}\!\! dt\, \langle \psi_- ,t|i\hbar\dt-\hat H| \psi_+ ,t\rangle=
W[I,J]-\int_{-\infty}^{+\infty}\!\! dt \left[J(t) \bq(t) + I(t) \bp(t)  \right]\, .
\ee
To prove that the values of $I$ and $J$ on the r.h.s. are the extremal ones it is necessary to take
derivatives of this equation with respect to the sources, and remember that on the l.h.s.
the extremal value cannot depend on the Lagrange multipliers, nor can the constraint points $\bp$ and $\bq$ on the r.h.s.

\section{Appendix C: The realization of the rising delta functional when $k\rightarrow \Lambda$.}

In order to analyze the $k\rightarrow \Lambda$ limit of  eq.~\eqref{intdiffeqk}
 we first perform a change of variables in the path integral:
$$ p^\prime=p-\bp+(r_k\dt)^{-1}\frac{\de \Gamma_k}{\de\bq}\quad , \quad
   q^\prime=q-\bq-(r_k\dt)^{-1}\frac{\de \Gamma_k}{\de\bp}      $$
and then define the complex variable: $z=(p^\prime-i q^\prime)/\sqrt{2}$.
The result of these manipulations is:
\ba
e^{i \Gamma_k[\bp,\bq]}=\int \left[ dz \right]\mu_k 
\exp i{\bigg\{\half \int\!\! dt\, \left(z^* r_k \id z - z r_k \id z^*\right)
-\frac{\de \Gamma_k}{\de\bq}\cdot(r_k\dt)^{-1}\frac{\de \Gamma_k}{\de\bp}}\nonumber\\
+S\left[\bp-(r_k\dt)^{-1}\frac{\de \Gamma_k}{\de\bq}+\sqrt{2}\Re(z),\bq+(r_k\dt)^{-1}\frac{\de H_k}{\de\bp}-\sqrt{2}\Im(z)\right]
\bigg\}\, . \nonumber
\ea
Under the assumption that $\Gamma_k$ stays finite for any $k\in [0,\Lambda]$, when $k\rightarrow \Lambda$
every $\Gamma_k$-dependent term on the right hand side (r.h.s.) gets killed by the divergence of
$r_k$. On the other hand, since $\mu_k={\rm Det}\left(\frac{1+r_k}{2\pi}\de\right)$ (excluding the possible zero eigenvalues),
 the first term in the exponent together
with the regularized functional measure provides a rising delta functional, constraining
$z$, i.e. $(p-\bp)$ and $(q-\bq)$, to vanish.\footnote{
Although the quadratic form $(r_k\id\de)$ in the exponent
and the operator in the measure $(1+r_k)\de$ asymptotically differ for a factor $\id$, 
the path integral is properly normalized~\cite{VZ}
in such a way to be finite for a free system ($\forall k \in [0,\Lambda]$)
and to show a $k$-independent divergence in the $H=0$ case.}
Thus in this limit the r.h.s. reduces to
$\exp\{i S[\bp,\bq]\}$ and the AEHA coincides with the bare Hamiltonian action.
To show that a rising delta functional is indeed realized we need to prove that the quadratic form
$\left(z^* r_k \id z - z r_k \id z^*\right)$ is positive definite. This is not obvious
since $\id$ is a real operator on the spaces of functions one is usually interested in, 
but whose sign is not fixed. 
However, if the domain of the functional integral is such that 
all contributions coming from the time boundaries are vanishing,
and if the Fourier transform is allowed, then one can write 
(the reader should interpret the integrals as generic sums over unspecified domains)
\ba
&&\frac{i}{2}\!\!\int_t  \left(z(t)^* r_k \id z(t) - z(t) r_k \id z(t)^*\right)
=\int_t p^\prime(t) O_k \id q^\prime(t)\nonumber\\
&=&\half\int_E r_k(E^2)E \left( p^\prime(-E) q^\prime(E)- q^\prime(-E) p^\prime(E) \right)\nonumber\\
&=&\int_E \theta(E)r_k(E^2)E \left( p^\prime(-E) q^\prime(E)- q^\prime(-E) p^\prime(E) \right)\nonumber\\
&=&i\int_E \theta(E)r_k(E^2)E \left( x_-(E)^* x_-(E)- x_+(E)^* x_+(E) \right)\nonumber
\ea
where we assumed $q(t)$ and $p(t)$ real, such that for their Fourier transforms
satisfy $p(-E)=p(E)^*$ and $q(-E)=q(E)^*$,
we defined $x_\pm(E)=\left( p^\prime(E)\pm i q^\prime(E)\right)/\sqrt{2}$, 
and we denoted by $\theta$ the Heaviside step function.
The last equation shows that the diagonalization of the quadratic form gives two complex Gaussians which
can be independently rotated to real Gaussians with positive definite inverse variances.
In reality they might be not positive definite and allow for zero modes, but
we will not discuss this possibility in the present work.

\section{Appendix D: The Average Effective Hamiltonian in Euclidean space and Wick rotation.}
Of course the Hamiltonian formalism without time makes little sense. 
However it could be nice to forget about the evaluation of integrals with poles once and for all
by working in Euclidean space from the very beginning.
In this appendix the reader will find the translation, of some of the main formulas of the present work
to Euclidean space and a discussion on the possible equivalence of the theories in Minkowski
and Euclidean space, i.e. on the feasibility of a Wick rotation to imaginary time.

Let's start with scalar QM. In this case Wick rotation ($t\rightarrow -i\tau$) of eq.~\eqref{hampathint1} with action~\eqref{QMhamiltonianaction} is safe and leads to a convergent path integral
\be
e^{W[I,J]}=\int \left[dp dq\right]\mu[p,q] 
e^{\,-\{S[p,q]-Jq-Ip\}} \nonumber 
\ee
with action 
\be
S[p,q]=\int d\tau\left[-p(\tau)i\partial_\tau q(\tau)+H(p(\tau),q(\tau))\right]\,.
\ee
The regularization goes as usual
\be
e^{W_k[I,J]}=\int \left[ dp dq\right]\mu_k e^{\,-\{S[p,q]+\Delta S_k[p,q]-Jq-Ip\}}\nonumber
\ee
with $\Delta S_k$ and $\mu_k$ which can still be chosen according to 
formulas~\eqref{deltaSk} to~\eqref{diagmu} if we replace $\dt$ with 
$-i\partial_\tau$ (the minus sign here is due to the global minus factorized in front of the action).
The definition of the AEHA is
\be
\Gamma_k\left[\bp,\bq\right] +\Delta S_k\left[\bp,\bq\right]=
\mathop{\rm ext}_{I,J}\left(I \cdot\bp +J\cdot \bq-W_k[I,J] \right)\nonumber
\ee
which is equivalent to
\be
e^{-\Gamma_k[\bp,\bq]}=\int \left[ dp dq\right]\mu_k[p,q]  
e^{\,-\left\{S[p,q]+\Delta S_k[p-\bp,q-\bq]
-(p-\bp)\frac{\de \Gamma_k}{\de \bp}-(q-\bq)\frac{\de \Gamma_k}{\de \bq}\right\} }\,.\\
\ee
From it the flow equation follows
\be
\dot{\Gamma}_k=
\half {\rm Tr}\left[\left(\Gamma_k^{(2)}+R_k\de\right)^{-1}\dot{R}_k\de\right]-\frac{\dot{\mu}_k}{\mu_k}
\ee
where $R_k\de=\Delta S_k^{(2)}$.
We see that this equation formally differs from the Minkowskian
one~\eqref{preflow2} by the absence of the imaginary factor $i$ on the l.h.s,
by a global minus factor on the r.h.s. and by the fact that inside
$R_k$ we find the operator $i\partial_\tau$ instead of $\dt$.
Thus, for instance, in the particular case of an off-diagonal regulator
the explicit form of the flow equation becomes
\ba
\dot{\Gamma}_k&=&
{\rm Tr}\left[(-\dot{r}_k i\partial\de)\left(
\left(-r_k i\partial\de+\frac{\de^2\Gamma_k}{\de \bq\de\bp}\right)
-\frac{\de^2\Gamma_k}{\de \bp\de\bp}
\left(r_k i\partial\de+\frac{\de^2\Gamma_k}{\de \bp\de\bq}\right)^{-1}
\frac{\de^2\Gamma_k}{\de \bq\de\bq}
\right)^{-1}
\right]\nonumber\\
&-&{\rm Tr}\left[\dot{r}_k\left(1+r_k\right)^{-1}\de\right]\,\,.
\ea

Next let's consider scalar covariant Hamiltonian QFT. Since $\pi^\nu$ is a vector, Wick rotation involves also its zero component, whether or not we allow for transverse momenta:
 $x^0\rightarrow-ix^4$ and $\pi^0\rightarrow-i\pi^4$.
However, performing such a replacement in the action~\eqref{covariant action}
 with Hamiltonian~\eqref{bareH_QFT} one finds that 
$$iS\rightarrow\int\!\!d^dx
\left[\half(\pi-\partial\phi)^2-\half(\partial\phi)^2-V(\phi)\right]$$
therefore the integral over $\pi$ diverges. 
In other words such a Wick rotation cannot be performed. 
The main difference from the case of QM, or the reason for such a failure,
is the fact that the momenta are assumed to rotate along with time.
Despite this problem, one possible reason for studying a
Euclidean covariant Hamiltonian formulation is that we know that
the Euclidean non-covariant Hamiltonian theory makes perfectly sense because it is  related by a continuos Wick rotation to the corresponding Minkowskian theory.
Therefore the Euclidean covariant formulation can be derived from the non-covariant Hamiltonian formulation and studied as a generalization of it.
By definition the bare action of such a covariant Hamiltonian Euclidean theory
reads
\be
S[\pi^\nu,\phi]=\int d^dx\left[-\pi^\nu i\partial_\nu \phi
+\cH(\pi^\nu,\phi)\right]\,.
\ee
Its Hamiltonian quantization in a scheme where only longitudinal momenta
are present is based on the functional integral
\be
Z[I_\nu,J]=\int \left[d\pi^\nu d\phi\right]\de\!\left[\Pi_\perp^{\nu\rho}\pi_\rho\right]
\mu\,
e^{\,-\left\{ S[\pi^\nu,\phi]-I_\nu\cdot\pi^\nu-J\cdot\phi\right\} } \, .\nonumber
\ee
Again, to get a functional RG flow equation representation
of this integral on introduces a $k$-dependence in the bare action and in the measure.
In the following we choose an off diagonal quadratic regularization, i.e.
of the kind~\eqref{deltaSkQFT} and~\eqref{mukQFT}, 
but with $\partial_\nu$ replaced by  $i\partial_\nu$.
The definition of the AEHA is the same as in Euclidean quantum mechanics
\be\label{defGammasQFT}
\Gamma_k\left[\bpi^\nu,\bphi\right] +\Delta S_k\left[\bpi^\nu,\bphi\right]=
\mathop{\rm ext}_{I_\nu,J}\left(I_\nu \cdot\bpi^\nu +J\cdot\bphi- W_k[I_\nu,J] \right)
\ee
wherefrom the usual integro-differential equation
\be
e^{- \Gamma_k[\bpi^\nu,\bphi]}=\int \left[ d\pi^\nu d\phi\right]
\de\!\left[\Pi_\perp^{\nu\rho}\pi_\rho\right]\mu_k \, 
e^{\, -\left\{S[\pi^\nu,\phi]+\Delta S_k[(\pi-\bpi)^\nu,\phi-\bphi]
-(\pi-\bpi)^\nu\frac{\de \Gamma_k}{\de \bpi^\nu}-(\phi-\bphi)\frac{\de \Gamma_k}{\de \bphi}\right\} }\,.\\
\ee
Again, the Euclidean flow equation can be obtained 
from the Minkowskian one by stripping the imaginary $i$ on the l.h.s.,
by changing the global sign on the r.h.s. and by replacing 
$r_k\partial\de$ with $r_k i\partial\de$.

As far as fermions are concerned, no new behavior under Wick rotation shows up, because
of the identification of configuration space with the reduced phase space.


\end{document}